\documentclass[acmtog, authorversion]{acmart}

\usepackage{graphicx}
\usepackage{soul}
\usepackage{dsfont}
\usepackage{gensymb}
\usepackage{microtype}
\usepackage{pifont}
\usepackage{threeparttable}

\AtBeginDocument{%
  \providecommand\BibTeX{{%
    \normalfont B\kern-0.5em{\scshape i\kern-0.25em b}\kern-0.8em\TeX}}}

\setcopyright{acmlicensed}
\acmJournal{TOG}
\acmYear{2021}
\acmVolume{40}
\acmNumber{6}
\acmArticle{224}
\acmMonth{12}
\acmDOI{10.1145/3478513.3480538}

\definecolor{darkgreen}{HTML}{3F7D31}
\definecolor{darkred}{HTML}{BA3132}

\newcommand{\cmark}{\textcolor{darkgreen}{\ding{51}}}
\newcommand{\xmark}{\textcolor{darkred}{\ding{55}}}

\newcommand{\eg}{e.g.,\ }
\newcommand{\ie}{i.e.,\ }
\newcommand{\etal}{et~al.\ }

\begin{document}

\title{FreeStyleGAN: Free-view Editable Portrait Rendering with the Camera Manifold}

\author{Thomas Leimk\"uhler}
\email{thomas.leimkuehler@mpi-inf.mpg.de}

\author{George Drettakis}
\affiliation{
	\institution{Universit\'{e} C\^{o}te d'Azur and Inria}
	\country{France}	
}
\email{george.drettakis@inria.fr}

\begin{abstract}
Current Generative Adversarial Networks (GANs) produce photorealistic renderings of portrait images.
Embedding real images into the latent space of such models enables high-level image editing.
While recent methods provide considerable semantic control over the (re-)generated images, they can only generate a limited set of viewpoints and cannot explicitly control the camera.
Such 3D camera control is required for 3D virtual and mixed reality applications.
In our solution, we use a few images of a face to perform 3D reconstruction, and
we introduce the notion of the GAN \emph{camera manifold}, the key element allowing us to precisely define the range of images that the GAN can reproduce in a stable manner.
We train a small face-specific neural implicit representation network to map a captured face to this manifold and complement it with a warping scheme to obtain free-viewpoint novel-view synthesis.
We show how our approach -- due to its precise camera control -- enables the integration of a pre-trained StyleGAN into standard 3D rendering pipelines, allowing e.g., stereo rendering or consistent insertion of faces in synthetic 3D environments.
Our solution proposes the first truly free-viewpoint rendering of realistic faces at interactive rates, using only a small number of casual photos as input, while simultaneously allowing semantic editing capabilities, such as facial expression or lighting changes.
\end{abstract}



\begin{teaserfigure}
	\includegraphics[width=\textwidth]{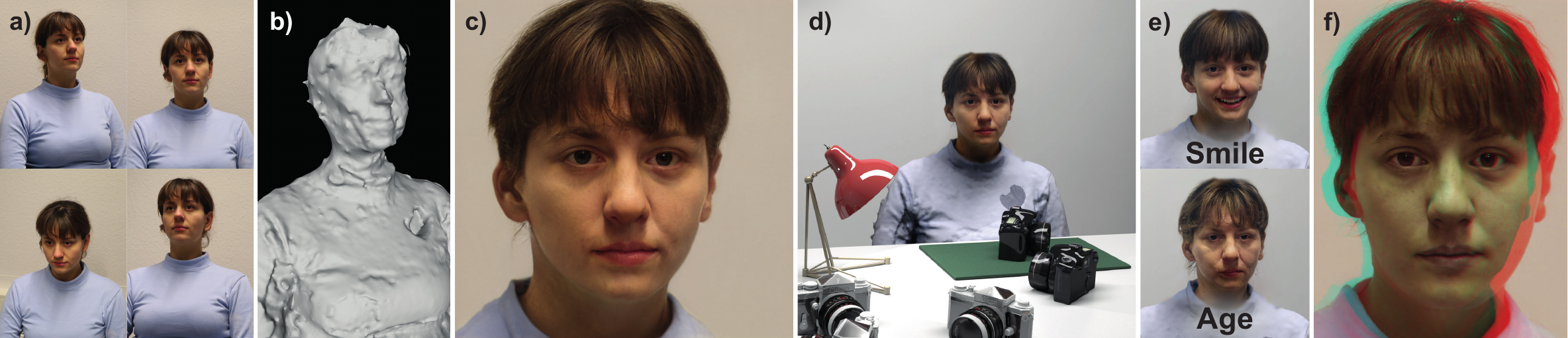}
	\caption{
	We introduce a new approach that generates an image with StyleGAN defined by a \emph{precise 3D camera}. This enables faces synthesized with StyleGAN to be used in 3D free-viewpoint rendering, while also allowing all semantic editing provided by GAN methods. 
    Our method takes as input multiple views of a person (examples in \emph{a}), used to reconstruct a coarse 3D mesh (\emph{b}).
    To render a novel view, we identify the closest camera which corresponds to an image that StyleGAN can generate (\emph{c}).
    We lift this view to 3D and obtain free-viewpoint renderings with arbitrary camera models, which allows the integration of our renderings into synthetic 3D scenes (\emph{d}).
     We inherit the high-quality semantic editing capabilities from StyleGAN (\emph{(e)} smile or aging), and enable stereoscopic rendering (\emph{f}).
    Our method can be integrated into any rendering pipeline and - for the first time - marries generative image modeling with traditional rendering.
}
\label{fig:teaser}
\end{teaserfigure}

\maketitle


\section{Introduction}

Recent Generative Adversarial Networks (GAN) can generate stunningly realistic images of faces from a latent code~\cite{karras2019style, karras2020analyzing}. Combined with recent methods to project a real photo to such a latent code~\cite{abdal2019image2stylegan, tewari2020pie} they also allow semantic image editing, e.g., controlling facial expression or relighting~\cite{harkonen2020ganspace, abdal2020styleflow, deng2020disentangled}. This opens the ground-breaking potential of replacing the expensive and complex process of capturing~\cite{debevec2000acquiring}, animating and rendering human faces \cite{seymour2017meet} by simply using GAN-generated images based on photos. Unfortunately, two central components are missing for this process to become reality: currently there is no way to consistently generate a realistic GAN image of a person in a Computer Graphics (CG) shot as defined by a \emph{precise and complete 3D camera} (3D position, 3D rotation, and field of view), and current methods can only generate a limited set of viewpoints. We present the first solution that solves these problems, allowing highly realistic, \emph{editable} faces to be correctly rendered for a any given 3D camera, and thus seamlessly integrated into standard CG imagery. Our method opens the way to using such highly realistic, editable face renderings for a wide range of applications such as virtual or augmented reality, immersive video-conferencing or even immersive remote training.

We build on the StyleGAN architecture \cite{karras2019style, karras2020analyzing} that is inherently a 2D image generator. 
Recent methods \cite{tewari2020stylerig, abdal2020styleflow, deng2020disentangled} can render head poses parameterized by two angles, but have no way to generate an image for a specific and complete 3D camera, ignoring at least five degrees of freedom (DoF).
This results in a small subspace of 3D camera poses; the nature and limits of this subspace have never been precisely defined, much less extended.

Our first challenge is thus to characterize the subspace of 3D camera parameters that StyleGAN can represent, enabling the design of a method for free viewpoint face rendering.
We analyze this subspace, carefully defining what we call the \emph{camera manifold}.
This definition is inherently linked to the alignment step performed on the face database used to train the StyleGAN portrait model, severely constraining the 2D locations of eyes and mouth. We also determine the boundaries of the camera manifold, allowing us to project a free-view camera onto it.

Providing full 3D capabilities for StyleGAN requires at least some 3D information; we use a few (10-25), casually captured photos of a face as input and generate coarse face geometry and calibrated cameras to guide part of our method.

Given these photos of a person, our second challenge is finding the latent code to provide to StyleGAN so it can generate the corresponding image. To do this, we need our camera manifold: we first find the closest view \emph{on the manifold}, then train a small, per-face latent representation network to find the latent vector for a given manifold view. We train the network with calibrated input views and image-based renderings using the reconstructed face geometry.

However, many free-viewpoint camera poses are not on the manifold. This raises our third and final challenge, \ie to provide a method allowing any such view to be synthesized. We do this by warping from the closest camera on the manifold to the desired novel view, using the coarse reconstructed geometry.

These three steps allow us to introduce a fully-operational, interactive system with \emph{precise 3D camera control} while fully exploiting StyleGAN-quality photorealistic face synthesis and corresponding manipulations (see Fig. \ref{fig:teaser}).
We consider a pre-trained and fixed StyleGAN2 model in this work, which allows us to run our per-face pre-processing pipeline within 45 minutes on a single GPU.

In summary our contributions are:
\begin{itemize}
\item An in-depth study and quantitative definition of the subspace of camera poses that StyleGAN can robustly synthesize, that we call \emph{camera manifold}.
\item A method to generate a realistic StyleGAN face based on a precisely defined 3D camera pose on the manifold.
\item A warping scheme to render any camera pose freely defined in 3D, fully consistent with semantic editing methods. 
\end{itemize}
Our system allows interactive synthesis of realistic faces, allowing free-viewpoint navigation in 3D while moving around a face casually captured with a handful of photos. We demonstrate interactive sessions with several captured faces, also performing semantic edits in a view-consistent manner -- such as changing facial expression, opening/closing eyes/mouth -- and also camera model manipulation (e.g., the Vertigo effect), seamless integration with synthetic 3D environments, and the first ever stereoscopic GAN renderings.


\section{Related Work}
\label{sec:relatedWork}

Our solution touches on several vast domains: Generative adversarial networks (GANs), image-based rendering (IBR), face models and portait rendering, and constrained camera models. In what follows, we only review the most closely related work to ours.

\subsection{Image Synthesis and Editing with GANs}
\label{sec:gans}

Generative adversarial networks~\cite{goodfellow2014generative} build a statistical model trained to mimic the distribution of training data (often images). In just a few years, GANs have evolved to produce truly photorealistic images at high resolutions.
Currently, StyleGAN \cite{karras2019style, karras2020analyzing, karras2020training} marks the state of the art in unconditional image generation in narrow domains such as faces, cars or cats. 
Importantly, a single forward pass through the generator provides photo-realistic imagery, enabling interactive rendering.

Latent codes represent semantically meaningful and reasonably disentangled concepts for many neural models~\cite{upchurch2017deep, radford2015unsupervised, karras2019style}. 
In StyleGAN, an informed manipulation of latents results in high-level image changes, \eg in pose, lighting, facial expression, age, gender, etc. 
The manipulations can correspond to linear \cite{harkonen2020ganspace, tewari2020stylerig, shen2020interpreting} or non-linear \cite{abdal2020styleflow, br2021photoapp} paths in latent space, while disentangled controls can be jointly trained with the generator \cite{deng2020disentangled}.
Our approach is fully compatible with semantic manipulation, such as the method of H\"ark\"onen \etal \shortcite{harkonen2020ganspace}, allowing us to dynamically edit our renderings.

All methods that rely on latent manipulations to change semantic attributes are necessarily bound to the span of the training data, especially when an alignment step is used \cite{jahanian2019steerability}. 
In practice, this means that only a subset of possible camera poses can be synthesized by StyleGAN.
Portrait alignment restricts cameras to 3\,DoF: two rotations and field of view. 
While previous work considers only the two rotations, ours is the first method that allows free view navigation with complete (7+\,DoF) camera models.

Recent work recovers latent codes from a \emph{single} image \cite{zhu2016generative, abdal2019image2stylegan, richardson2020encoding} enabling semantic editing of real photos with unprecedented quality \cite{abdal2020image2stylegan++, tewari2020pie}.
In contrast we are the first to devise a parameterized embedding of multiple views of the same face, which requires a mapping from 3D cameras to latents.

The StyleGAN latent space can be used to represent \emph{any} image with high fidelity -- even ones far outside the training data distribution~\citep{abdal2019image2stylegan} -- at the cost of low-quality edits.
Consequently, researchers have investigated regularizations to enforce latents closer to the original distribution.
This has been done by considering the distribution of the latents directly 
\cite{wulff2020improving, zhu2020indomain, tewari2020pie}, or by utilizing semantic knowledge about the images generated \cite{richardson2020encoding}. We use a prior as well to find a good compromise between photorealism and identity preservation.

In recent work, GANs have been used as a multi-view generator to aid inverse rendering \cite{zhang2021image}, or to estimate 3D cameras, shape, and lighting \cite{pan20202d, shi2021lifting}.
A different line of work has devised GANs to incorporate 3D information directly, e.g., using
3D geometry~\cite{zhu2018visual} or 3D semantic occupancy~\cite{chen2020free} followed by image-to-image translation for final output, or using volumetric generator layers followed by projection for rotations and scaling~\cite{nguyen2019hologan}. GRAF \cite{schwarz2020graf} and pi-GAN \cite{chan2020pi} produce a 3D radiance field, which can then be rendered using volume rendering. 
Such GANs with explicit 3D-awareness are promising, but so far fail to generate high-quality results.  In contrast to these solutions, we lift a pre-trained 2D GAN to 3D, and thus inherit the advantages of 2D GANs (high quality, high resolution, robust training procedures, etc.) while enhancing the method with precise 3D camera control.
3D-aware generative modeling can be improved by jointly learning the distribution of camera parameters \cite{niemeyer2021campari}.
Similarly, we estimate this distribution for a pre-trained StyleGAN model to refine our camera manifold.


\subsection{Multi-view Free-viewpoint Image-based Rendering}

Traditional IBR usually relies on explicit proxy geometry from structure-from-motion/multi-view stereo \cite{schonberger2016structure}:
Unstructured lumigraph rendering (ULR) \cite{buehler2001unstructured} uses a geometric proxy to heuristically blend multi-view images. We use this method for training. Recently, \emph{neural rendering} has made significant advances in the quality of novel view synthesis~\cite{Tewari2020NeuralSTAR}.
Geometry-based methods use deep learning to improve the quality of novel view synthesis~\cite{hedman2018deep,riegler2020free}, but are restricted to static scenes.
Single-object mesh-based neural rendering approaches \cite{thies2019neural, thies2019image} also use proxy geometry for reprojection followed by neural refinement. 
We share the methodology of using a geometric proxy, and -- like these solutions -- enable true free-viewpoint rendering.
However, we are the first to exploit geometry to steer a GAN for free-viewpoint face rendering.

Neural Volumes~\cite{lombardi2019neural} learn a volumetric scene representation, which is rendered using ray marching.
They can handle video sequences, but require synchronized and calibrated cameras.
Flexible neural scene representations allow high-level scene editing \cite{li2020crowdsampling}, but they are restricted to changes captured in the input views. In contrast, our approach exploits the expressive space of variations captured in a generative model.

Recent solutions include neural radiance fields \cite{mildenhall2020nerf, zhang2020nerf}, that use implicit scene representations \cite{sitzmann2019srns} and volume rendering techniques for view synthesis. Extensions allow deforming objects \cite{park2020nerfies} or changing lighting \cite{nerv2021, martinbrualla2020nerfw}, offering some degree of semantic control \cite{zhang2021editable}.
These methods have long rendering times (usually tens of seconds per frame) and most of them need dense capture; in contrast, we exploit the power of GANs allowing interactive rendering and use of sparse capture (10-25 images).


\subsection{Face Models and Portrait Rendering}

Manually created face rigs \cite{seymour2017meet} are stunningly realistic, but require skilled visual effects artists working long hours. 3D morphable models (3DMMs) \cite{egger20203d,blanz1999morphable} are generative models that allow fine-grained control over shape and texture of an object, often used for faces.
Such models can be augmented with deformation-dependent texture maps \cite{matthews2004active}
and photometric capture to also model appearance variation \cite{smith2020morphable}.
3DMMs offer maximal control over shape, facial expression, texture, etc., but tend to lack photo-realism and typically only model the face region (hair, neck, etc. are not included) even though recent advances \cite{yenamandra2020i3dmm} are starting to consider entire heads.

Another approach to capturing faces involves complex multi-view setups \cite{beeler2010high,ghosh2011multiview, lombardi2018deep, wei2019vr, bi2021deep}.
In contrast to these methods, we only need a few casually captured images as input and can perform arbitrary semantic edits.

Face re-enactment is an active area of research.
Kim \etal \shortcite{kim2018deep} transfer head pose and facial expressions from a source video to target video. 
Single \cite{geng2018warp, siarohin2020first, zakharov20} or multiple \cite{wang2019fewshotvid2vid, zakharov2019few} source images can be used in conjunction with facial feature point extraction of a target sequence to hallucinate novel views and facial expressions. 
Most of these methods have only been demonstrated with very limited pose variation.
Our approach allows true free-viewpoint navigation while reasonably preserving identiy.

Rao \etal \shortcite{rao2020free} shares some similarities with our approach, since they also process facial animations in a canonical frame, and lift the face to 3D. However, they use StyleGAN to synthesize only the mouth, and require dense capture.

Face rotation using a single image \cite{nagano2019deep, xu2020deep, zhou2020rotate} usually proceeds by first estimating geometry, followed by a neural rendering step. While results are impressive given that only a single image is used, the variation of attainable poses tends to be limited.
Subtle perspective corrections of portraits -- including stereo image synthesis -- can be addressed using image warping \cite{fried2016perspective, zhao2019learning}.
We also use warping, but mostly for global transformations, while perspective effects are generated by manipulating latent codes, successfully overcoming typical problems arising from disocclusions, hair, and the lack of global consistency.

Specialized variants of neural radiance fields for portrait synthesis \cite{wang2020learning, gafni2020dynamic} produce free-viewpoint results of high visual quality and can handle facial animations, but inherit the problems of neural radiance fields: dense capture setup and long rendering times.
Gao \etal \shortcite{gao2020portraitnerf} combine neural radiance fields with strong priors to allow portrait rendering from a single view, achieving some effects we handle (e.g., the Vertigo-effect), but the range of achievable poses and visual quality are limited.


\subsection{Constrained Cameras}

Camera control in interactive applications is important for viewpoint computation, motion planning and editing. See Christie \etal \shortcite{christie2008camera} for an in-depth survey.
Our camera manifold builds on fixed projected locations, related to 2-object composition problems \cite{blinn1988where}: How to place a camera such that two objects are at prescribed locations in the image plane?
Lino and Christie \shortcite{lino2012efficient} observed that the solution to this problem is a manifold surface in camera parameter space; they later generalized this to the \emph{Toric space} \cite{lino2015intuitive}, allowing interactive exploration and optimization of viewpoints with multiple geometric constraints.
Similarly, we parameterize the 7\,DoF camera space using a 3D manifold. However, their approach is more general, with a more complex solution space, while our simpler problem has more constraints and is amenable to efficient convex optimization.


\section{Overview}
\label{sec:overview}

\begin{figure}
	\includegraphics[width=0.99\linewidth]{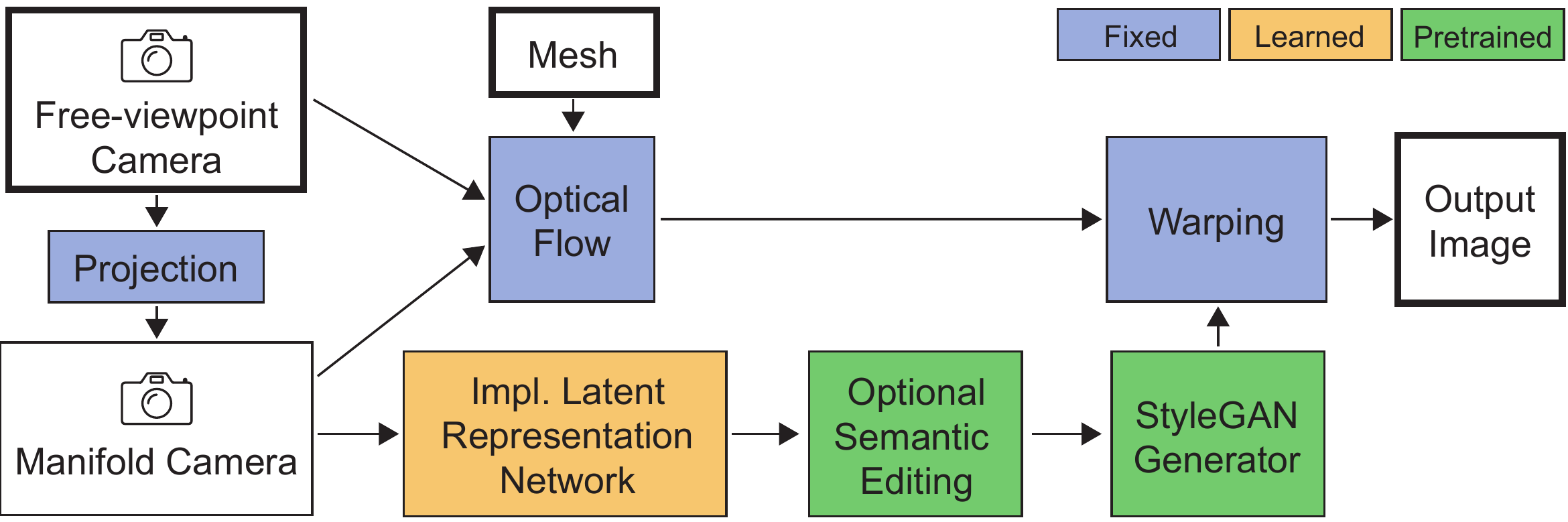}
	\vspace*{-2mm}
	\caption{
	Overview of our method.
	}
	\label{fig:overview}
\end{figure}

Our method takes as input 10-25 photographs of a person and allows free 3D viewpoint rendering of the face with arbitrary camera models based on a pre-trained StyleGAN2 portrait model \cite{karras2020analyzing}.
An overview of our method is shown in Fig.~\ref{fig:overview}.

\subsection{Method}

Our key observation is that the StyleGAN portrait model can synthesize a quantifiable range of pose variations:
All generated images have the property that the 2D locations of eyes and mouth are severely constrained.
Based on this observation, we define a space of camera parameters we call the \emph{camera manifold}, which results in images in the canonical configuration that respects this constraint (Sec.~\ref{sec:camera_manifold}).
We then analyze and delimit the \emph{range} of the manifold, \ie the subspace of camera poses \emph{within} our manifold parameterization that StyleGAN can synthesize successfully.

We start from a few photos and use standard 3D reconstruction to obtain calibrated cameras and mesh of the face, as described below.
To render the face from an unconstrained novel view $\textbf{V}$, we use the steps described next.

We first project $\textbf{V}$ to the closest camera $\hat{\textbf{V}}$ on the manifold, thereby reducing the problem to an in-domain rendering task.
We then use an implicit latent representation network, which maps physically meaningful coordinates on the camera manifold to a StyleGAN latent code (Sec.~\ref{sec:preprocess}).
The network functions as a parameterized embedding of the face and is trained in a supervised fashion using aligned input views and simple IBR renderings \cite{buehler2001unstructured} in a progressive training schedule.
The entire pre-processing pipeline is face-specific and takes 45 minutes.

The final step is to move from the StyleGAN-generated manifold view $\hat{\textbf{V}}$ to free view $\textbf{V}$ (Sec.~\ref{sec:runtime}).
Given that both views correspond to physically meaningful cameras, we use the face mesh to establish a dense flow field which describes how the image corresponding to the manifold view $\hat{\textbf{V}}$ needs to be deformed to obtain the image corresponding to (unconstrained) novel view $\textbf{V}$.

\subsection{Preprocessing}
\label{sec:calibration}

To incorporate 3D awareness, our method relies on calibration of the input cameras and an approximate geometric representation of the face to be rendered. Neither calibration nor geometry need to be accurate for our method to produce convincing free-viewpoint images.

We calibrate cameras and create the geometric proxy using off-the-shelf software \cite{realitycapture}, resulting in  
a reconstructed triangle mesh, the calibrated cameras and undistorted input images (see supplemental for details). Quality is satisfactory using 10-25 photos, despite casual sequential capture without a rig.


\section{The Camera Manifold}
\label{sec:camera_manifold}

To allow free-viewpoint rendering of captured faces, we first need to understand the native capabilities of StyleGAN in terms of viewpoint synthesis.
To this end, we define the subset of cameras that allows the generation of valid StyleGAN2 images.
The portrait model published by the authors \cite{karras2020analyzing} is trained on the FFHQ dataset \cite{karras2019style}. The alignment of this dataset is a key component of the model, greatly improving quality but limiting the variety of images that can be synthesized. This is because the photos lie on a relatively narrow image manifold \cite{jahanian2019steerability}, implicitly defined by this alignment process based on facial features.
We define our \emph{camera manifold} based on the FFHQ alignment procedure (Sec.~\ref{sec:manifold_definition}), estimate the range of the manifold in which StyleGAN can generate realistic images (Sec.~\ref{sec:manifold_range}), explain how to project a free-viewpoint camera to the closest manifold camera (Sec.~\ref{sec:manifold_projection}), and finally present the process of aligning a captured face to the canonical coordinate system we use (Sec.~\ref{sec:alignment}).

\begin{figure}[!h]
	\includegraphics[width=0.99\linewidth]{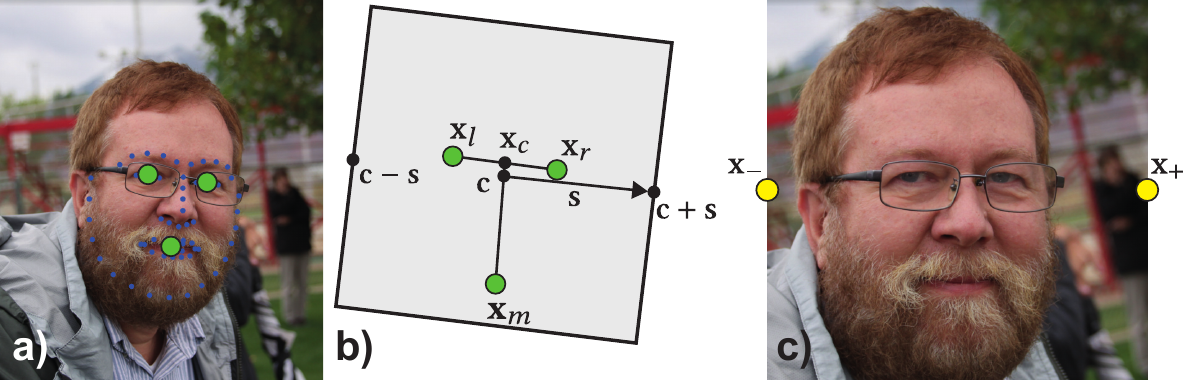}
	\vspace*{-2mm}
	\caption{
		The 2D alignment performed in the FFHQ dataset.
		\emph{a)} Raw facial feature points (blue dots) are detected and aggregated to obtain representative eye and mouth positions (green dots).
		\emph{b)} Geometric features are used to determine the square crop window (grey, not shown to scale) with center $\mathbf{c}$ and vector $\mathbf{s}$ giving orientation and size.		
		\emph{c)} The resulting aligned image.
	\vspace*{-2mm}
		}
	\label{fig:ffhq_alignment}
\end{figure}

The FFHQ dataset was constructed by first collecting images from Flickr, followed by a cleanup step and alignment, where first 68 facial features are found \cite{kazemi2014one} (blue dots in Fig.~\ref{fig:ffhq_alignment}a). 
Then the eye and mouth features are aggregated to obtain representative eye positions $\mathbf{x}_l$ and $\mathbf{x}_r$, as well as a mouth position $\mathbf{x}_m$ in the image (green dots in Fig.~\ref{fig:ffhq_alignment}a).
From these three points a square crop window is computed by
determining its center $\mathbf{c}$ (also using the eye midpoint $\mathbf{x}_c$) as well as its orientation and size $\mathbf{s}$, all illustrated in Fig.~\ref{fig:ffhq_alignment}b. Exact formulas are given in the supplemental.
Given the crop window geometry, the original image is re-sampled to obtain the final aligned output image 
(Fig.~\ref{fig:ffhq_alignment}c).

\begin{figure*}[!h]
	\includegraphics[width=0.85\linewidth]{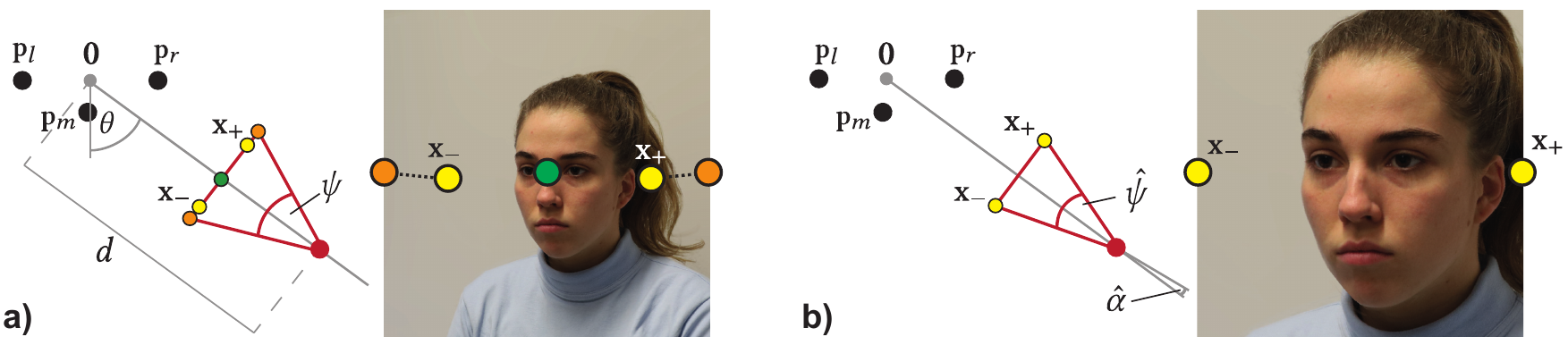}
	\vspace*{-2mm}
	\caption{
	The construction of the camera manifold.
	2D illustrations capture the setup from the top.
	\emph{a)} Our setup consists of 3D eye and mouth positions $\mathbf{p}_l$, $\mathbf{p}_r$, and $\mathbf{p}_m$, and a perspective camera (red) parameterized by spherical coordinates (only $\theta$ and $d$ are shown in this 2D illustration).
	The trackball parameterization in Eq.~\ref{eq:trackballParam} keeps the eye midpoint (green point) fixed to the image center, but fails to move $\mathbf{x}_-$ and $\mathbf{x}_+$ (yellow points) to their 2D target positions (orange points). 
	\emph{b)} Our manifold camera model (Eq.~\ref{eq:manifoldCamera}) optimizes for an additional 3D camera rotation (only $\hat{\alpha}$ is shown in the 2D illustration) and the field of view $\hat{\psi}$ to ensure that projected locations are fixed to their targets. The resulting image lies in the span of the StyleGAN portrait model.
	}
	\label{fig:manifold}
\end{figure*}

\subsection{Defining the Camera Manifold}
\label{sec:manifold_definition}

The alignment procedure described above guides the definition of our \emph{camera manifold}.
We first write the alignment as a mapping from facial feature positions to a 2D similarity transform:
$
\mathbf{F}(\mathbf{x}_l, \mathbf{x}_r, \mathbf{x}_m)
\in
\mathds{R}^6 \rightarrow (\mathds{R}^2 \rightarrow \mathds{R}^2).
$
The similarity transform maps pixel locations from the unaligned to the aligned image.
We observe that $\mathbf{F}$ is deterministic and $C^0$-continuous.
Since $\mathbf{F}$ is an alignment procedure, it exhibits an infinite number of invariances.
In our model, we require two invariances for an unambiguous mapping and we consider the two (arbitrary) points (Fig.~\ref{fig:ffhq_alignment}b)
\begin{equation*}
\mathbf{x_-}
=
\left[ 0, 0.5 \right]^T
=
\mathbf{F}(\mathbf{x}_l, \mathbf{x}_r, \mathbf{x}_m)
(\mathbf{c} - \mathbf{s}) 
=:
\mathbf{F}_-(\mathbf{x}_l, \mathbf{x}_r, \mathbf{x}_m)
\end{equation*}
and
\begin{equation*}
\mathbf{x_+}
=
\left[ 1, 0.5 \right]^T
=
\mathbf{F}(\mathbf{x}_l, \mathbf{x}_r, \mathbf{x}_m)
(\mathbf{c} + \mathbf{s}) 
=:
\mathbf{F}_+(\mathbf{x}_l, \mathbf{x}_r, \mathbf{x}_m),
\end{equation*}
where the definitions of $\mathbf{F}_-$ and $\mathbf{F}_+$ $\in \mathds{R}^6 \rightarrow \mathds{R}^2$  exploit the fact that $\mathbf{c}$ and $\mathbf{s}$ are themselves functions of the facial feature positions.

Our method relates the fixed positions $\mathbf{x_-}$ and $\mathbf{x_+}$ to feasible camera parameters, thus connecting 2D image alignment to 3D-aware image formation.
To map a 3D camera pose to aligned image space, we solve the following problem:
Assuming fixed eye and mouth positions in \emph{3D space}, which combinations of intrinsic and extrinsic camera parameters yield the prescribed projection locations $\mathbf{x_-}$ and $\mathbf{x_+}$ in image space?

For the purpose of defining our camera manifold, we assume that StyleGAN-generated images are modeled as being captured with a perspective pinhole camera. Note however, that we can generate images using arbitrary camera models.
We further assume, without loss of generality, that the left and right 3D eye positions are
\begin{eqnarray}
\label{eq:canon}
\mathbf{p}_l = \left[ -1, 0, 0 \right]^T
\quad
\text{and}
\quad 
\mathbf{p}_r = \left[ 1, 0, 0 \right]^T,
\end{eqnarray}
respectively.
We further assume we have access to the 3D mouth position $\mathbf{p}_m$ (see supplemental for our exact definition of a frontal pose).
In Sec.~\ref{sec:alignment} we show how to align a general pose of a face capture to this canonical setting.
Note that this alignment does not impose any restrictions on face shape.

A full camera model with 7\,DoF gives too much freedom for our manifold, as many camera parameters correspond to an image with the face lying outside the frame. We therefore, as a first step, restrict ourselves to a trackball camera model:
\begin{equation}
\label{eq:trackballParam}
\bar{\textbf{A}}(\theta, \phi, d, \psi)
=
\textbf{P}(\psi)
\textbf{T}(0,0,d)
\textbf{R}(\theta, \phi, 0),
\end{equation}
where
$\textbf{R} \in \mathds{R}^3 \rightarrow \mathds{R}^3$ is a rotation parameterized by three Euler angles,
$\textbf{T} \in \mathds{R}^3 \rightarrow \mathds{R}^3$ is a translation parameterized by three offset coordinates,
and
$\textbf{P} \in \mathds{R}^3 \rightarrow \mathds{R}^2$ is a perspective projection with field of view $\psi$.
The free parameters $\theta$, $\phi$, $d$, and $\psi$ correspond to horizontal and vertical rotations around the eye midpoint, distance to the eye midpoint, and field of view, respectively (Fig.~\ref{fig:manifold}a).
$\bar{\textbf{A}}$ corresponds to an un-aligned image in the spirit of Fig.~\ref{fig:ffhq_alignment}a:
The face lies within the frame, but is not aligned in general.

In a second step, we enable fixed projection locations as follows:
\begin{equation}
\label{eq:manifoldCamera}
\hat{\textbf{A}}(\theta, \phi, d)
=
\textbf{P}(\hat{\psi})
\textbf{R}(\hat{\alpha}, \hat{\beta}, \hat{\gamma})
\textbf{T}(0,0,d)
\textbf{R}(\theta, \phi, 0).
\end{equation}
Here, we add an additional rotation before the projection and exclude $\psi$ from the list of free parameters.
We refer to the free parameters
$
\mathbf{m} 
= 
\left[ \theta, \phi, d \right]^T
\in \mathds{M} \subset \mathds{R}^3
$ 
as \emph{manifold coordinates}.
We define the space of $\mathds{M}$ in Sec.~\ref{sec:manifold_range}.
In terms of the camera position, $\mathbf{m}$ corresponds to spherical coordinates.
Intuitively, the coefficients $\hat{\mathbf{c}} = \left[ \hat{\alpha}, \hat{\beta}, \hat{\gamma}, \hat{\psi}\right]$ steer a rotation (via $\hat{\alpha}$, $\hat{\beta}$, $\hat{\gamma}$)  and scaling (via $\hat{\psi}$)  without changing the camera position.
Excluding degenerate cases, which we avoid as described in Sec.~\ref{sec:manifold_range}, for any fixed set of manifold coordinates, there exists a coefficient vector 
$\hat{\mathbf{c}}(\mathbf{m})$ to position two 3D points to arbitrary 2D projected image locations. 

We now have all the machinery in place to solve our alignment problem (Fig.~\ref{fig:manifold}b), i.e., given a manifold coordinate $\mathbf{m}$, find a coefficient vector $\hat{\mathbf{c}}(\mathbf{m})$ that simultaneously satisfies the four equations:

\begin{align}
\label{eq:manifoldEquations}
\begin{cases}
\mathbf{F}_-
\left(
\hat{\textbf{A}}(\mathbf{m})\mathbf{p}_l,
\hat{\textbf{A}}(\mathbf{m})\mathbf{p}_r, 
\hat{\textbf{A}}(\mathbf{m})\mathbf{p}_m 
\right)
=
\mathbf{x}_-
\\ 
\mathbf{F}_+
\left(
\hat{\textbf{A}}(\mathbf{m})\mathbf{p}_l,
\hat{\textbf{A}}(\mathbf{m})\mathbf{p}_r, 
\hat{\textbf{A}}(\mathbf{m})\mathbf{p}_m 
\right)
=
\mathbf{x}_+ 
\\ 
\end{cases} 
\end{align}

\noindent
where $\hat{\textbf{A}}(\mathbf{m})\mathbf{p}$ denotes the projection of 3D point $\mathbf{p}$ to the screen using camera $\hat{\textbf{A}}(\mathbf{m})$.
Due to the nonlinear nature of the equations, obtaining an analytic solution is challenging.
Instead, we solve numerically for the least-squares solution using the Levenberg-Marquardt algorithm, which usually converges after 35 iterations.
The optimization takes about 10 milliseconds in our Python implementation and produces temporally stable results.
We therefore solve the equations on the fly.

Our model allows free positioning of the camera via $\mathbf{m}$.
However, all rotational DoF as well as the field of view are automatically adjusted to ensure aligned portrait images.
Our cameras therefore lie on a 3D manifold in the 7D parameter space.
The coordinates $\theta$ and $\phi$ resemble yaw/pitch parameterizations of face pose in previous work, but by construction also include translational and scaling components which previous approaches do not model.
Manipulation of $d$ results in the Vertigo effect: Increasing $d$, \ie moving the camera away from the face, is compensated by a corresponding decrease in the field of view, and vice versa.

\subsection{Range}
\label{sec:manifold_range}

The above definition does not restrict camera position $\mathbf{m}$, allowing arbitrary views of the head, \eg profile or rear views.
We observe, however, that StyleGAN-generated images have smaller variations in camera pose, consistently showing front or moderate oblique angle views.
We next determine the boundaries or \emph{range} of our camera manifold.
This mainly concerns the rotation parameters $\theta$ and $\phi$; the Vertigo-type variations induced by $d$ are small enough to ignore in this context, when avoiding very small $d$ (and resulting large $\hat{\psi}$).
Therefore, we simply enforce $d \geq 10$, \ie camera positions at least five inter-ocular distances away from the face.

\begin{figure}
	\includegraphics[width=0.99\linewidth]{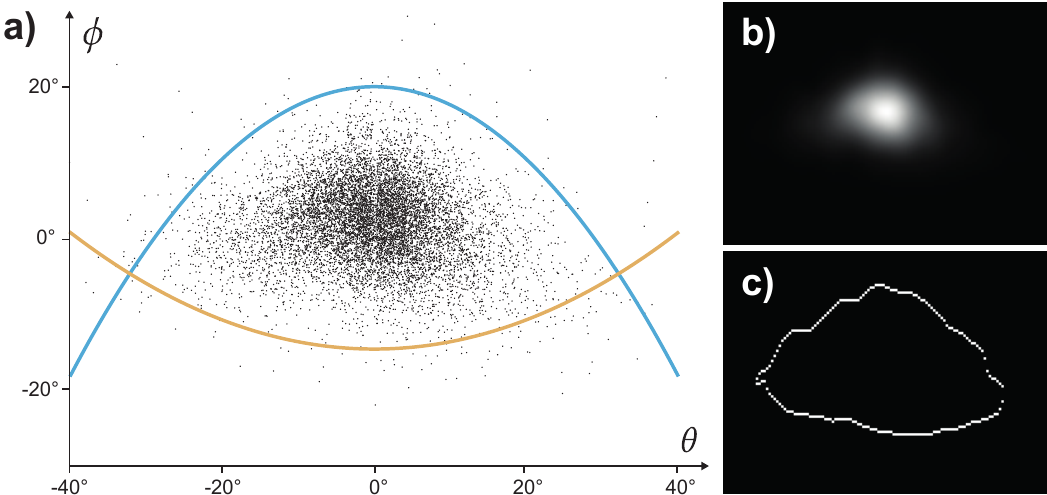}
	\vspace*{-2mm}
	\caption{
	Camera manifold range analysis.
	Sampling and analyzing StyleGAN-generated images in terms of camera pose (dots in \emph{a}) reveals an anisotropic distribution, which we bound using two parabolas (curves in \emph{a}).
	Bounding proceeds by first converting the samples into a probability distribution (\emph{b}), followed by iso-line extraction (\emph{c}) and curve fitting.
	\vspace*{-2mm}
		}
	\label{fig:manifold_range}
\end{figure}

To quantify the range of camera positions in the $\theta\phi$-plane, we first produce 10k random facial images using the StyleGAN2 generator with default truncation parameter $\psi=0.5$.
We then use the method of Bulat and Tzimiropoulos \shortcite{bulat2017far} to infer estimates of 3D eye and mouth positions per image, which we convert to corresponding $\theta\phi$-tuples (details on calibration in supplemental).
When plotting the resulting distribution (dots in Fig.~\ref{fig:manifold_range}a), we see that the images indeed concentrate around the frontal pose $\theta = \phi = 0$, but show a strong eye-shaped anisotropy, while being close to symmetric around the axis $\theta=0$.
These observations suggest that we can model the boundary of this distribution using two parabolas of the form
$\phi = a \theta^2 + b$.
To obtain the coefficients $a$ and $b$, we first convert the samples into a probability distribution $p_{\theta\phi}$ (Fig.~\ref{fig:manifold_range}b).
We use kernel density estimation on a regular grid of size 
$128 \times 96$ 
capturing 
$\theta \in [-40\degree, 40\degree]$
and  
$\phi \in [-30\degree, 30\degree]$,
utilizing a Gaussian kernel with a standard deviation of three grid cells.
We then determine the iso-lines
$p_{\theta\phi}=0.01 \max(p_{\theta\phi})$
(Fig.~\ref{fig:manifold_range}c) and perform two least-squares parabola fits.
Below, we give the result of our fit for the upper and lower boundary curves (parabolas in Fig.~\ref{fig:manifold_range}a):
\[
c_u(\theta)=-0.024 \theta^2 + 20.00
\quad
\text{and}
\quad 
c_l(\theta)=0.010 \theta^2 - 14.60.
\]
A valid coordinate in our model satisfies
$c_l(\theta) \leq \phi \leq c_u(\theta)$.

\subsection{Projecting a Camera to the Manifold}
\label{sec:manifold_projection}

\begin{figure}
	\includegraphics[width=0.7\linewidth]{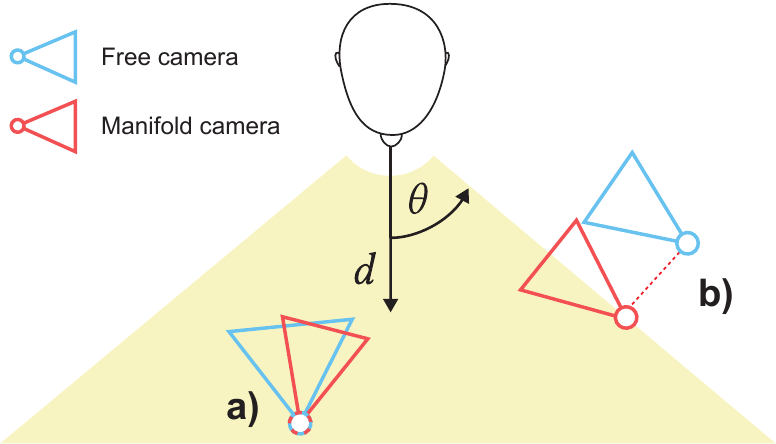}
	\vspace*{-2mm}
	\caption{
	Projecting cameras to the camera manifold (only $\theta$ and $d$ are shown in this top view):
	\emph{a}) If the position of the free-viewpoint camera (blue) is in the valid range (orange area, corresponds to a $\phi$-slice of Fig.~\ref{fig:manifold_range}a), the closest manifold camera (pink) only differs by in-place rotation and field of view.
	\emph{b}) If the free-viewpoint camera lies outside the valid range, the camera position is also affected by the manifold projection. Consequently, parallax occurs only in case \emph{b)}.
	\vspace*{-3mm}
		}
	\label{fig:manifold_projection}
\end{figure}

Finding the closest camera on the manifold for a free-viewpoint camera is easy in our model (Fig.~\ref{fig:manifold_projection}):
We take the camera position in spherical coordinates as manifold coordinate $\mathbf{m}$ and solve Eq.~\ref{eq:manifoldEquations}.

However, aribitrary camera positions do not necessarily correspond to valid manifold coordinates as defined in the previous section. 
In these cases, we find the closest valid manifold coordinate (Fig.~\ref{fig:manifold_projection}b), using an
analytic solution for this projection to the valid range (details in supplemental).
Note that only in this case the camera position changes and parallax is induced in the image.

\subsection{Alignment to Canonical Coordinates}
\label{sec:alignment}

Our head reconstruction can have arbitrary global 3D scene scale, positioning and orientation.
Since our camera manifold assumes the eyes to be at defined 3D positions $\mathbf{p}_l$ and $\mathbf{p}_r$, we perform 3D alignment to the canonical configuration (Eq.~\ref{eq:canon}) \cite{rao2020free, gao2020portraitnerf}, making our method independent of the reconstruction algorithm. 

The locations of the eyes and mouth on the mesh are required for alignment. 
While 3D facial feature location algorithms exist \cite{bowyer2006survey}, our multi-view data provided stable results despite possibly strong reconstruction noise:
Similar to the FFHQ alignment, we use the method of Kezemi and Sullivan \shortcite{kazemi2014one} to obtain facial feature points on the face for each input view and compute representative 2D eye and mouth locations (Fig.~\ref{fig:ffhq_alignment}a).
We then re-project the 2D feature points onto the mesh and average the re-projected features of all input views to obtain a robust estimate of the eye and the mouth positions in 3D.

We now seek to find a similarity transform that maps the estimated 3D eye positions to their targets $\mathbf{p}_l$ and $\mathbf{p}_r$.
To obtain a unique solution and remain consistent with the definition of the camera manifold, we further enforce a frontal pose in terms of vertical orientation, as detailed in the supplemental.
The transformation matrix is obtained using the Levenberg-Marquardt algorithm and then used to transform both the mesh and the input camera parameters.

We also process the input images:
We observe that the embedding of images into the GAN latent space (Sec.~\ref{sec:preprocess}) gives results of higher visual quality when the background is smooth. 
Further, such ``simple'' backgrounds are easier to integrate 
into a synthetic scene.
We therefore blur the background of the input images.
We use the method of Ke \etal \shortcite{ke2020green} to extract a foreground matte.
Then we apply a strong Gaussian filter (we use a $\sigma$ of one tenth the image width) to the background region \cite{knutsson1993normalized} and compose the foreground on top.
Finally, we 2D-align all input images using the FFHQ alignment.

Recall, that in the canonical coordinate system, any camera's manifold coordinate $\mathbf{m}$ corresponds to the camera's position in spherical coordinates.
Consequently, the manifold coordinates of the 3D-aligned input cameras -- provided they are in the valid range -- correspond to the 2D-aligned images in the sense of our model.


\section{Mapping Manifold Coordinates \\to StyleGAN Latent Codes}
\label{sec:preprocess}

We now have the manifold coordinates that correspond to a given camera pose.
We next find the StyleGAN latent code that corresponds to a given manifold coordinate, by training a small per-face implicit latent representation network.
This network allows us to move on the camera manifold.

\subsection{StyleGAN Terminology}

StyleGAN~\cite{karras2019style} maps normally distributed random samples $\mathbf{z} \in \mathds{R}^{512}$ to an intermediate latent vector $\mathbf{w} \in \mathds{R}^{512}$ using a learned mapping $\mathbf{w} = H(\mathbf{z})$. The space of $\mathbf{w}$'s is commonly referred to as $W$.
The vector $\mathbf{w}$ controls feature normalizations in 18 layers of the generator network $G$, which produces the final image $I = G(\mathbf{w}) = G(H(\mathbf{z}))$.
It has been observed \cite{karras2019style}\cite{abdal2019image2stylegan} that the expressivity of the generator is much higher, when different $\mathbf{w}$ are fed to the generator layers.
In the general case, 18 different sets of latents $\mathbf{w^+} \in \mathds{R}^{18 \times 512} = W^+$ can be used. 

\subsection{Method}

By construction, all views $\hat{\mathbf{V}}$ on the camera manifold can be rendered by finding a corresponding latent vector 
$\mathbf{w}_{\textbf{m}} \in W^+$.
We therefore seek a mapping 
$M \in \mathds{M} \rightarrow W^+$ 
from manifold coordinates to latents, such that, given manifold coordinates $\textbf{m}$, we obtain 
$\hat{\mathbf{V}}$:
\[
\hat{\mathbf{V}}
=
G
\left(
\mathbf{w}_{\textbf{m}}
\right)
=
G
\left(
M\left(\mathbf{m}\right)\right).
\]
Recall that, in contrast to all previous methods, $\mathbf{m}$ is an exact meaningful physical 3D quantity.

We found a face-specific mapping $M$ to provide highest-quality results.
Therefore, in analogy to recent works on implict neural representations \cite{sitzmann2019srns, genova2019learning, sitzmann2020implicit}, we refer to $M$ as an implicit latent representation network.
It has been observed that the latents fed to the earlier StyleGAN layers correspond to coarse-scale image properties including face pose, while later layers add medium- to small-scale features to the images \cite{karras2019style}.
We therefore define a constrained architecture:
For layers 0-5 we parameterize $M$ with a small multi-layer perceptron (MLP) per layer, each mapping raw manifold coordinates $\mathbf{m}$ to a 512-D latent vector.
Due to the low-frequency behaviour of pose changes, we do not use Fourier features \cite{tancik2020fourfeat}.
For the MLPs, we found two hidden layers with 32 features and leaky ReLU activation functions to be sufficient for our application.
For the remaining layers 6-17 we directly optimize for static latents, which ensures a consistent output for different $\mathbf{m}$ (Fig.~\ref{fig:training}).

\subsection{Training Data}

Thanks to our multi-view setup and the geometric reconstruction, we can train $M$ in a supervised fashion using training tuples
$
\left\{
\left(
\mathbf{m}_k,
\hat{\mathbf{V}}_k
\right)
\right\}_k
$.
We employ two complementary sources of information as training data:
First, we use the aligned input images from Sec.~\ref{sec:alignment}. 
The images constitute a sparse set of high-quality training samples on the camera manifold.
Second, we use the input views together with the face mesh to create manifold views using image-based rendering.
Specifically, we use unstructured lumigraph rendering \cite{buehler2001unstructured} with cameras corresponding to random $\mathbf{m}$ on the camera manifold.
We can generate any of these renderings on the fly covering the entire manifold.
In the supplemental we give details on how we sample the valid manifold range.
The geometric reconstruction and calibration have uncertainty, producing rendering artifacts that reduce overall image quality.
We can compute a per-pixel estimate of this uncertainty by computing the color variance $\boldsymbol{\sigma}^2_c$ when blending the different images; we will use this estimate to reduce the influence of incorrect IBR pixels.

\subsection{Loss}

Our renderings should:
\emph{a}) be close to the training images,
\emph{b}) preserve the identity of the depicted person from all views, and
\emph{c}) look photo-realistic.
We use the following loss to achieve these goals:

\begin{equation}
\mathcal{L}
=
\mathcal{L}_{\ell_1}
+
\lambda_\text{LPIPS}
\mathcal{L}_\text{LPIPS}
+
\lambda_\text{id}
\mathcal{L}_\text{id}
+
\lambda_\text{prior}
\mathcal{L}_\text{prior}.
\end{equation}
The first two terms address goal \emph{a}).
Here, $\mathcal{L}_{\ell_1}$ penalizes pixel differences in terms of the $\ell_1$-norm, weighted by per-pixel IBR confidence:
\begin{equation}
\label{eq:l1Loss}
\mathcal{L}_{\ell_1}
=
\left\|
\exp
\left(
-\eta \boldsymbol{\sigma}^2_c
\right)
\left(
\hat{\textbf{V}}
-
\hat{\textbf{V}}_k
\right)
\right\|_1,
\end{equation}
with the falloff factor $\eta=100$ in all our experiments and $\boldsymbol{\sigma}^2_c = \mathbf{0}$ if $\hat{\textbf{V}}_k$ is an input view.
$\mathcal{L}_\text{LPIPS}$ is the LPIPS \cite{zhang2018unreasonable} distance which we apply to a downsampled version of the images to a resolution of $256 \times 256$ pixels.

The identity loss $\mathcal{L}_\text{id}$ addresses goal \emph{b}) and makes use of the pre-trained VGG-face \cite{parkhi2015deep} network $\Psi$, which converts an image into face recognition features:
\begin{equation*}
\mathcal{L}_\text{id}
=
1 - 
\langle
\Psi(\hat{\textbf{V}}),
\overline\Psi
\rangle,
\end{equation*}
where $\overline\Psi$ denotes the normalized mean face recognition features of all input views.

Finally, we address goal \emph{c}) by recognizing that StyleGAN renderings of highest quality and stability are usually obtained when the latents $\mathbf{w}$ follow the distribution dictated by the mapping network $H$.
Specifically, we address the problem that the extended space $W^+$, where the styles of different StyleGAN layers are decoupled from each other, leads to an underconstrained problem and consequently images of lower realism.
We enforce a certain amount of coherence between the styles using
\begin{equation*}
\mathcal{L}_\text{prior}
=
\frac{1}{18}
\sum_{i=1}^{18} 
\left\|
\mathbf{w}_i - \mathbf{\overline{w}}
\right\|_1,
\end{equation*}
where $\mathbf{w}_i$ refers to the latent fed into the $i$-th generator layer, and 
$\mathbf{\overline{w}}=\frac{1}{18} \sum_{i=1}^{18} \mathbf{w}_i$.
This prior encourages the network to produce latents which are similar for all generator layers, \ie closer to the original StyleGAN latent space $W$.

\begin{figure}
	\includegraphics[width=0.99\linewidth]{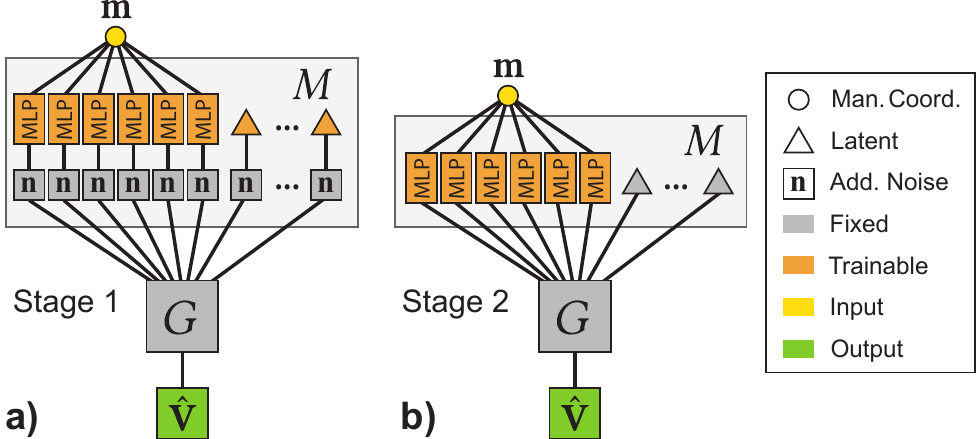}
	\vspace*{-2mm}
	\caption{
	We employ a progressive training schedule.
		\emph{a)} In the first stage, we only use the input images as training data. We train MLPs that map manifold coordinates (yellow dot) to the first 6 StyleGAN latents, and we directly optimize for the remaining static latents (orange triangles). All latents fed to $G$ are subject to random perturbations (boxes labelled $\mathbf{n}$) during training.
		\emph{b)} In the second stage, we fine-tune the MLPs by augmenting the training data with IBR and fix the static latents (grey triangles).
	\vspace*{-2mm}
		}
	\label{fig:training}
\end{figure}

\subsection{Training}
\label{sec:training}
We found that a progressive training schedule, which splits training into two stages, produces results of highest quality. 
Fig. \ref{fig:training} summarizes our method; details and exact parameters are given in supplemental.

In the first stage, we only use the aligned input views as training data and optimize all trainable parameters (Fig.~\ref{fig:training}a).
Intuitively, this training stage provides sparse anchors for the MLP, which is responsible for pose changes and at the same time optimizes the latents of the static GAN layers with the highest-possible quality training data. This stage trains in 35 min on an NVIDIA RTX6000.

In the second stage, we provide samples from the entire manifold as training data using a mixture of ULR renderings ($85\%$) and input views ($15\%$).
We fix the latents of the static layers to prevent high-frequency IBR artifacts from impacting them (Fig.~\ref{fig:training}b).
This stage fills in the pose gaps between the input views and trains in 4 min.


\section{Free-viewpoint Rendering}
\label{sec:runtime}

We have so far established a method to render a view  $\hat{\mathbf{V}}$ corresponding to a pinhole camera $\hat{\textbf{A}}$ on the camera manifold using StyleGAN.
We now seek to move away from the manifold to synthesize free-viewpoint images $\mathbf{V}$ with an arbitrary camera model $\textbf{A}$.
We achieve this goal using a simple image warping strategy as follows.

\begin{figure}[!h]
	\includegraphics[width=0.99\linewidth]{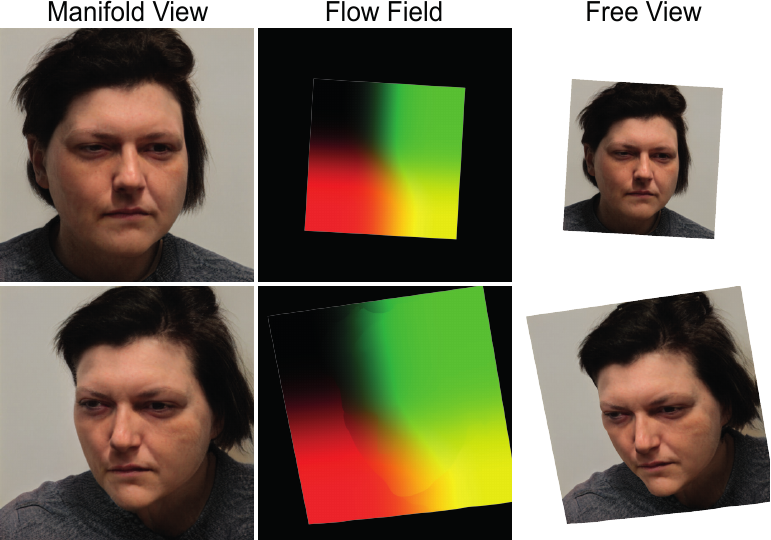}
	\vspace*{-1mm}
	\caption{
		Free-view generation in our approach:
		We use StyleGAN to generate the closest view on the camera manifold (left), which is warped using a flow field (center) to obtain 
		the final result (right).
		The top row shows an example of a parallax-free flow field, arising 			from a free camera in the valid manifold range.
		The configuration in the bottom row requires parallax in the flow field, as the free camera lies outside the valid manifold range.
\vspace{-2mm}
		}
	\label{fig:warping}
\end{figure}

First we project $\textbf{A}$ to the closest manifold camera $\hat{\textbf{A}}$ (Sec.~\ref{sec:manifold_projection}, first column in Fig.~\ref{fig:warping}).
Recall that both $\textbf{A}$ and $\hat{\textbf{A}}$ correspond to physically meaningful cameras.
Therefore, since we have the face geometry at our disposal, we can render an inverse flow field \cite{mark1997post, yang2011image}, which for each pixel of $\mathbf{V}$ indicates where to lookup $\hat{\mathbf{V}}$ (second column in Fig.~\ref{fig:warping}).
The desired viewpoint $\mathbf{V}$ is then obtained by warping $\hat{\mathbf{V}}$ according to the flow field (third column in Fig.~\ref{fig:warping}).
This procedure is compatible with any camera model $\mathbf{A}$ for $\mathbf{V}$, including physical lenses, stereoscopic setups, etc.
Multi-sampled effects require multiple entries per flow field pixel \cite{yu2010real}. 
We use multi-sampling by default to anti-alias occlusion boundaries arising from parallax.

\begin{figure*}[th]
	\includegraphics[width=0.99\linewidth]{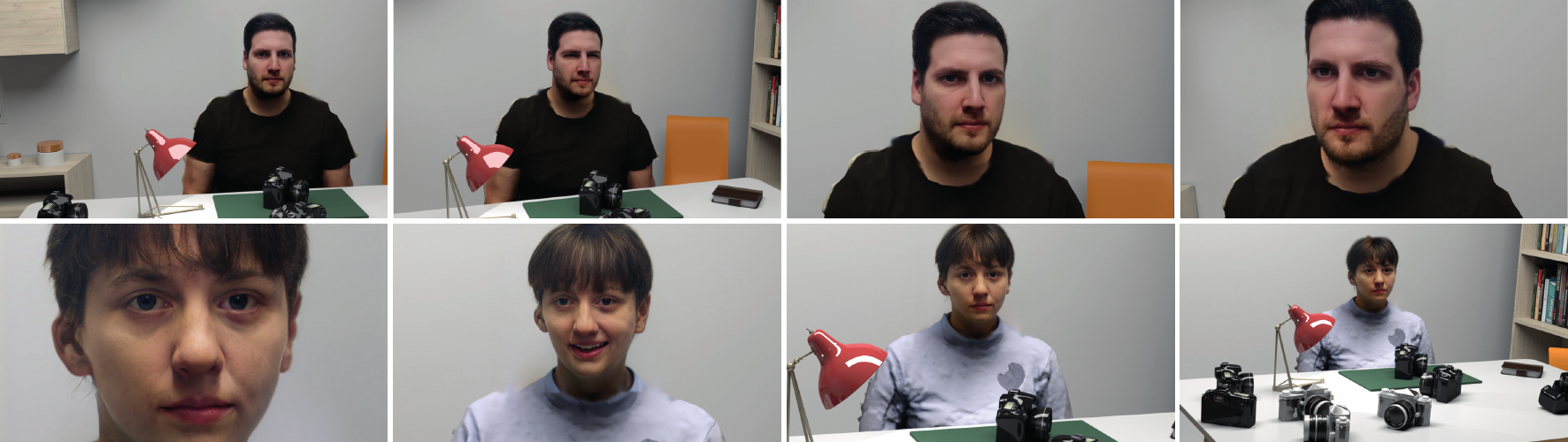}				\vspace*{-2mm}
	\caption{
		Compositing of our portrait renderings into a synthetic scene with a moving camera. Note the semantic edits (eyes, smile) in the second column.}
	\label{fig:result_synthetic}
\end{figure*}

Note that if the position of $\textbf{A}$ corresponds to a manifold coordinate $\mathbf{m}$ in the valid range (Sec.~\ref{sec:manifold_range}), the projection to $\hat{\textbf{A}}$ does not change the camera's position (Fig.~\ref{fig:manifold_projection}a).
Therefore, the flow field is parallax-free and corresponds to a continuous remapping (top row in Fig.~\ref{fig:warping}).
If $\textbf{A}$ is a perspective pinhole camera, the warp even reduces to a simple global affine image transformation.
Parallax occurs in the flow field only if $\textbf{A}$ moves outside the valid manfold range (Fig.~\ref{fig:manifold_projection}b; bottom row in Fig.~\ref{fig:warping}). 
This operation is akin to view-dependent texture mapping~\cite{debevec1996modeling}, with StyleGAN as a texture generator.

The final image $\mathbf{V}$ is always synthesized using a coordinated interplay between latent manipulations and image warping.
Even a seemingly simple lateral camera motion cannot be created by only shifting the image: The new viewpoint results in a slight variation of viewing angle that in turn leads to a change of manifold coordinates. A re-evaluation of the networks for synthesizing correct perspective is therefore required.


\section{Results and Evaluation}
\label{sec:evaluation}


We implemented our method based on the original StyleGAN2 code \cite{karras2020analyzing} in Python, complemented by a custom OpenGL rendering framework, including an interactive viewer. We provide all source code of our method here: \textcolor{blue}{\url{https://repo-sam.inria.fr/fungraph/freestylegan}}.

To acquire faces we requested authorization from our institutional ethics committee, and due to restrictions imposed by their decision concerning privacy and security, we are only able to show a small number of images per subject as illustration in this paper, but cannot distribute the full datasets.
We provide instructions on how to capture a face with a smartphone in the supplemental.

\begin{figure}[!h]
	\includegraphics[width=0.99\linewidth]{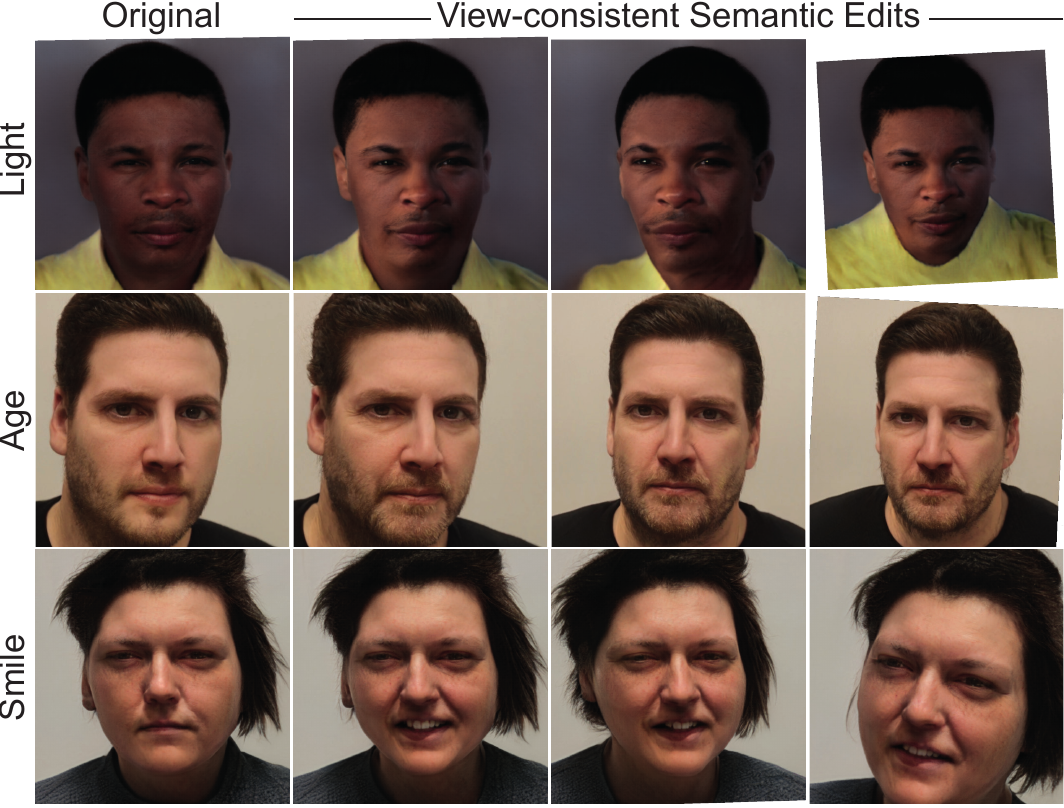}
	\vspace*{-2mm}
	\caption{
		Semantic editing results:
		The original multi-view embedding (col. 1) is modified (col. 2) and stays consistent across novel views (col. 3+4).
	\vspace*{-3mm}
}
	\label{fig:result_semantic_editing}
\end{figure}

\subsection{Results}
\label{sec:results}

To evaluate our approach, we use our own captures.
We asked participants for a series of 10-25 photographs of their head and upper body.
The participants were asked to sit still, while non-professional photographers with no background in visual computing sequentially captured images from different viewing angles, using a digital (smartphone) camera. 
None of the authors (or their collaborators) were present during capture and no training was conducted.

To ensure both successful geometry reconstruction and disentanglement of the camera pose in the StyleGAN latents, we found it crucial to eliminate all other sources of variation (lighting, facial expression, etc.) across the input views. Yet, our technique is robust enough to allow lightweight sequential capture and does not require a multi-camera setup.

To demonstrate free-viewpoint rendering and the accuracy of our camera model, we composite our results on top of a rendering of a synthetic scene in Fig.~\ref{fig:result_synthetic} with a moving camera.
We perform 
compositing with color correction membranes \cite{farbman2009coordinates} using convolution pyramids \cite{farbman2011convolution}.
As can also be seen from our supplemental video, our approach for the first time allows to successfully integrate a 2D GAN portrait rendering into a synthetic 3D scene with free-viewpoint control.

In Fig.~\ref{fig:result_semantic_editing} we show semantic editing results.
We apply the method of H\"ark\"onen \etal \shortcite{harkonen2020ganspace} to the output latents from our implicit representation network.
We observe that the modifications stay view-consistent in our renderings (see supplemental video for interactive editing sessions).
Note that the method used to perform semantic editing is independent of our approach.

\begin{figure}[!h]
	\includegraphics[width=0.99\linewidth]{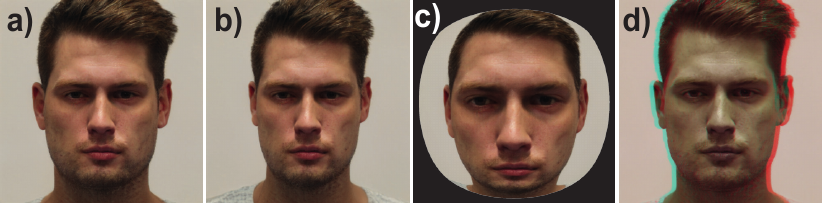}
	\vspace*{-2mm}
	\caption{
Different camera models:
Perspective pinhole camera (\emph{a}), 
Vertigo-effect (\emph{b}), non-linear lenses (\emph{c}), and stereo image pairs (\emph{d}, use anaglyph glasses for stereo 3D impression).
	\vspace*{-1mm}
}
	\label{fig:result_camera_models}
\end{figure}

In Fig.~\ref{fig:result_camera_models} we demonstrate face renderings with different camera models.
The Vertigo effect (Fig.~\ref{fig:result_camera_models}b) can be achieved without leaving the camera manifold, \ie without warping, by simply varying the $d$-component of the manifold coordinate.
We see that our latent representation network is able to synthesize the shift in perspective when moving towards the face while opening the field of view.
In Fig.~\ref{fig:result_camera_models}c we demonstrate spherical lens distortions, while Fig.~\ref{fig:result_camera_models}d showcases a stereoscopic rendering result.
Our solution generates the two images from the correct viewpoints, also allowing fine-grained control over interocular distance and screen depth.

\begin{figure}[!h]
	\includegraphics[width=0.99\linewidth]{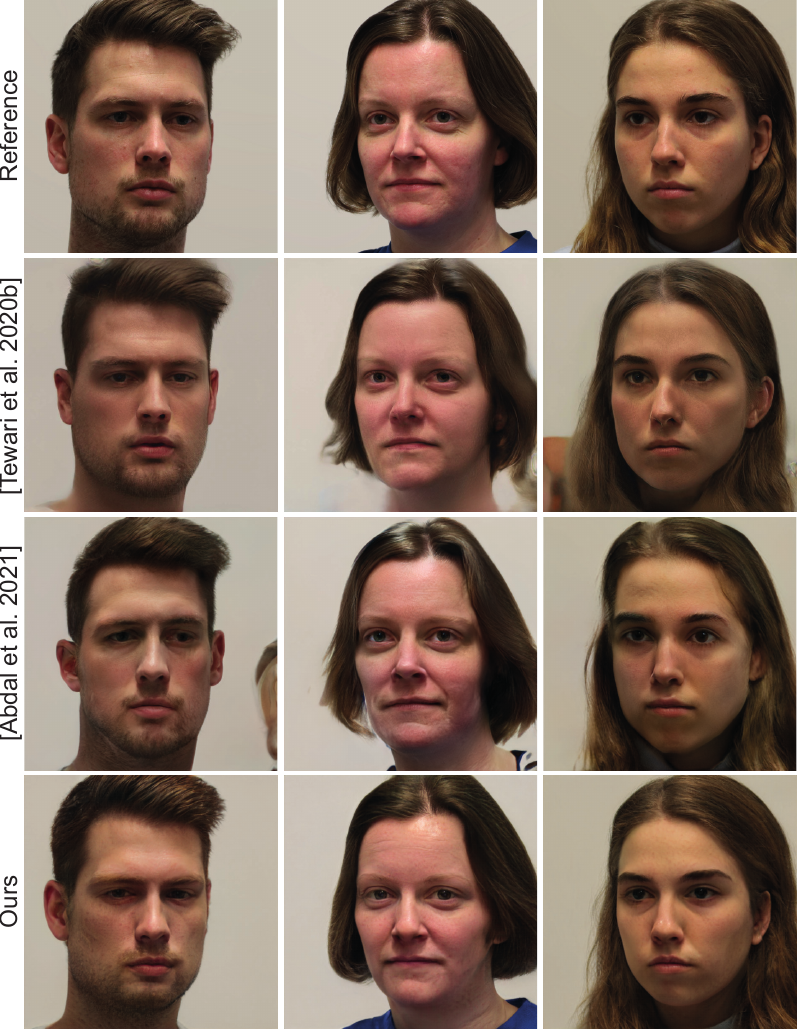}
	\vspace*{-2mm}
	\caption{
		Comparison to state-of-the-art StyleGAN editing approaches.
}
	\vspace{-2mm}
	\label{fig:comparison_stylegan}
\end{figure}

\begin{figure}[!h]
	\includegraphics[width=0.99\linewidth]{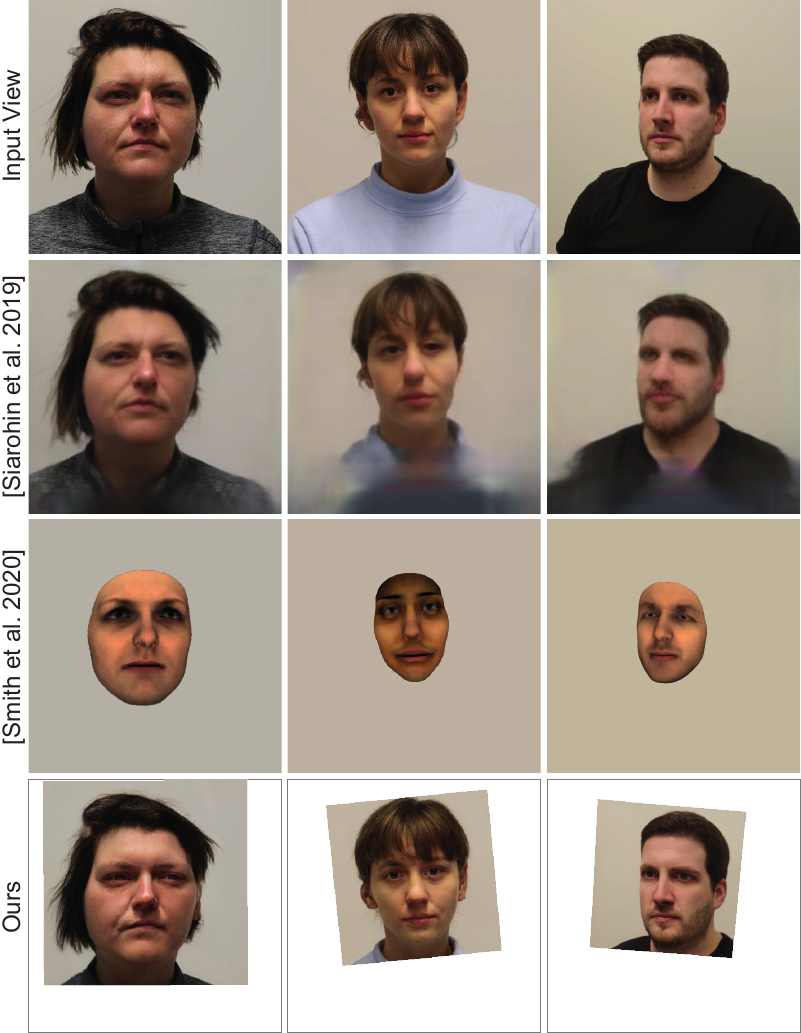}
	\vspace*{-2mm}
	\caption{
		Comparison to editable portrait rendering methods.}
	\vspace*{-4mm}
	\label{fig:comparisons_portrait}
\end{figure}

\subsection{Comparisons}
\label{sec:comparisons}

We compare our algorithm to state-of-the-art StyleGAN-based rendering techniques, portrait-specific approaches, and for completeness to general-purpose free-view IBR methods.
A full set of videos can be found in our supplemental material.

First we compare our approach to two state-of-the-art StyleGAN-based portrait rendering approaches: PIE \cite{tewari2020pie} and StyleFlow \cite{abdal2020styleflow}. While both methods perform novel-view portrait rendering using StyleGAN, they differ from ours in two fundamental ways:
First, they rely solely on manipulations of StyleGAN latent code $\mathbf{w}^+$, restricting novel views to the camera manifold, \ie a very small fraction of what our method allows, prohibiting free-viewpoint navigation.
Second, their view parameterization does not correspond to a physically meaningful quantity: 
Latent code-induced viewpoint manipulations are not simply a camera rotation around a fixed 3D point (Sec.~\ref{sec:camera_manifold}), but induce non-linear dependencies between camera position, orientation and field of view.
This is in contrast to our camera manifold formulation which translates latent codes into physically meaningful quantities.
To allow a fair comparison, we thus restrict our output to the camera manifold, omitting our free-viewpoint warping. We determine image quality by re-synthesizing held-out input views.
For both PIE and StyleFlow, we use a frontal view as input. Since there is no way to directly find the held-out camera with these methods, we densely sample the yaw-pitch-space for both methods and select the view minimizing mean $\ell_2$-distance of facial landmarks \cite{kazemi2014one} to the requested input view.
Our approach generates the required views directly.
Our method outperforms the others both numerically (Tbl.~\ref{tab:quant_stylegan_quality}) and, arguably, in subjective image quality (Fig.~\ref{fig:comparison_stylegan}).

\begin{table}[]
	\centering
	\caption{
	Image quality comparison of our method against state-of-the-art StyleGAN-based approaches.}
	\vspace*{-2mm}
	\label{tab:quant_stylegan_quality}
	\begin{tabular}{lrrr}
		Method & PSNR$\uparrow$ & SSIM$\uparrow$ & E-LPIPS$\downarrow$\\
		\toprule
		\cite{tewari2020pie} & 17.8 & .711 & .027  \\
		\cite{abdal2020styleflow} & 17.2 & .717 & .028  \\
		Ours & 20.8 & .758 & .020 \\
		\bottomrule
	\end{tabular}
	\vspace*{-2mm}
\end{table}

We also compare against the facial re-enactment technique of Siarohin \etal \shortcite{siarohin2020first} and the morphable face albedo model of Smith \etal \shortcite{smith2020morphable} in Fig.~\ref{fig:comparisons_portrait}.
For Siarohin \etal we use their pre-trained model based on the VoxCeleb dataset 
\cite{nagrani2017voxceleb}, which allows a reasonable variety of poses due to a conservative crop window.
We use a textured mesh \cite{realitycapture} rendered from free viewpoints (unless stated otherwise) to drive an aligned frontal-pose view.
We observe that their result quality highly depends on the viewpoint:
While frame-filling portrait views are handled well, viewpoint-induced scaling of the head tends to result in distortions or severe identity shifts.
Additionally, their spatial output resolution is $256\times256$, in contrast to our resolution of $1024\times1024$.
For the method of Smith \etal we use their inverse rendering pipeline based on a frontal view.
Their method is designed for maximum control over face shape, facial expression, and surface properties far beyond what the StyleGAN latent space captures, but it suffers from a lack of photo-realism.

\begin{figure}[!h]
	\includegraphics[width=0.99\linewidth]{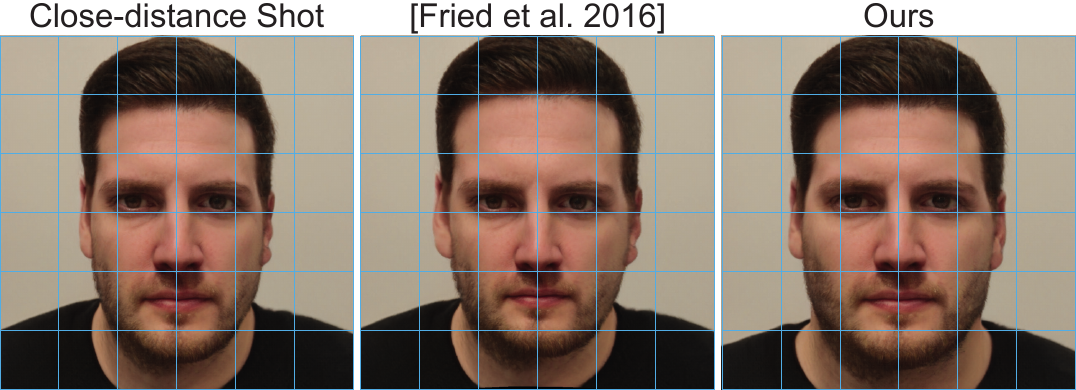}
	\vspace*{-2mm}
	\caption{
		Manipulation of camera distance and focal length (Vertigo effect) using a warping technique (center) and our latent-driven approach (right). Use the grid overlay to assess changes of proportions.
}
	\vspace{-2mm}
	\label{fig:comparison_warping}
\end{figure}

Our technique allows to change the distance between face and camera by manipulating StyleGAN latent codes using our camera manifold formulation and the trained network $M$.
In Fig.~\ref{fig:comparison_warping}, we compare this latent-driven shift of perspective to the single-image approach of Fried \etal \shortcite{fried2016perspective}.
Given a shot captured from a close distance (Fig.~\ref{fig:comparison_warping}, left), their method uses warping to simulate a long-distance shot with the corresponding decrease in field of view (Fig.~\ref{fig:comparison_warping}, center).
In contrast, our method (Fig.~\ref{fig:comparison_warping}, right) requires only a manipulation of manifold coordinates and no warping to synthesize this Vertigo effect.
We observe that, while both techniques exhibit similar characteristics in the central face region, our method produces globally more consistent results, \eg neck and shoulders are also correctly influenced by the shift of perspective.

We quantitatively evaluate camera accuracy, image quality, and identity preservation in Tbl.~\ref{tab:quant_free}.
For completeness, we include the multi-view IBR methods DeepBlending \cite{hedman2018deep}, Free View Synthesis \cite{riegler2020free}, and NeRF++ \cite{zhang2020nerf}.
For the first two approaches, we use the pre-trained models from the authors, which have not been trained on faces.
Note that, in contrast to ours, none of these methods allows semantic editing.

To estimate camera accuracy, we re-synthesize all input views for six subjects resulting in 132 images total, while holding out the input views to compare against for the approaches of Hedman \etal \shortcite{hedman2018deep}, Riegler and Koltun \shortcite{riegler2020free}, and ours.
For the method of Siarohin \etal \shortcite{siarohin2020first} we use the input views as the driving source.
We then determine the $\ell_2$-distance of 2D facial landmarks \cite{kazemi2014one} to those of the ground truth images.
To compensate for different image resolutions, we normalize this error by the estimated interocular distance per view.
We additionally report the success rate of the landmark detector, providing an indication of how realistic the generated faces are \cite{tewari2020pie}. 
To analyze the capability of identity preservation, we employ the method of Schroff \etal \shortcite{schroff2015facenet} to extract face recognition features for 280 free-viewpoint video frames across different subjects and measure the cosine distance to the normalized mean recognition features of the respective input views (last column in Tbl.~\ref{tab:quant_free}).
In the supplemental we also present visual comparisons and image error metrics. Since most free-viewpoint IBR methods reproject images they achieve better quality; recall however that they \emph{do not allow any semantic editing}.
In contrast, the method of Siarohin et~al., which allows semantic editing in the form of facial expressions, does not perform well in the free-viewpoint setting for the metrics we considered.

\begin{table}[]
\small
\begin{threeparttable}
	\centering
	\caption{
	Comparison of camera accuracy (measured by facial landmarks: Alignment \& Detection Rate), and face Recognition Error.}
	\label{tab:quant_free}
	\begin{tabular}{lcrrr}
		Method & Semantic & \multicolumn{2}{c}{Facial Landmarks} & Recog. \\		
		 & Editing & Align.$\downarrow$ & Det. Rate$\uparrow$ & Error$\downarrow$ \\
		\toprule
		\cite{hedman2018deep} & \xmark & .023 & 99\% & .07 \\
		\cite{riegler2020free} & \xmark & .027 & 100\% & .08 \\
		\cite{zhang2020nerf}\tnote{1} & \xmark & .018 & 100\% & .24   \\
		\cite{siarohin2020first} & \cmark\tnote{2} & .254 & 42\% & .23  \\				
		Ours & \cmark & .068 & 100\% & .14  \\
		\bottomrule
	\end{tabular}
\begin{tablenotes}\footnotesize
\item[1] Due to time constraints, we did not train a separate model for each leave-one-out image set, but only one model using all images per subject.
\item[2] Editing is restricted to facial animations using a driving video.
\end{tablenotes}
\end{threeparttable}
\end{table}


\subsection{Ablations}
\label{sec:ablations}

We analyze the effectiveness of individual components of our algorithm by ablation.
We consider distribution and number of input views, alternatives to our camera manifold formulation, loss terms and training procedure, and our background blur approach.

\subsubsection{Input Views}

\begin{figure}
	\includegraphics[width=0.99\linewidth]{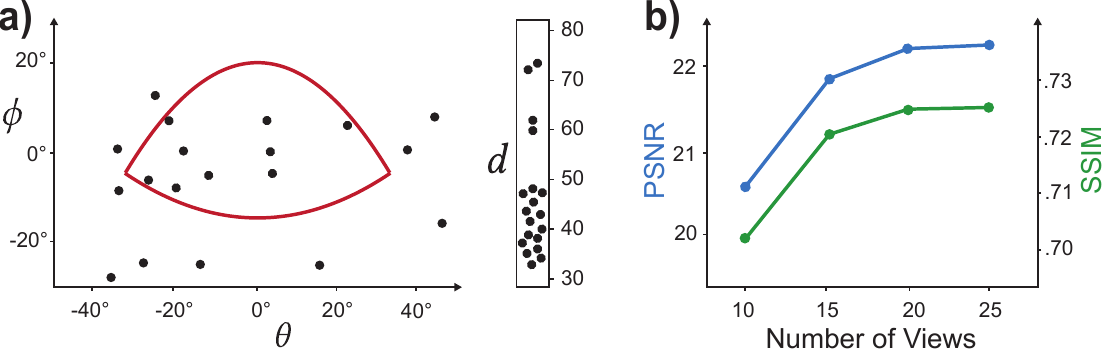}
	\caption{
		\emph{a)} A typical input view distribution (black dots) in our manifold parameterization. A large fraction of views lies outside the manifold boundary (red shape).
		\emph{b)} Image quality as a function of input view count.
	\vspace*{-2mm}
	}
	\label{fig:analysis_cameras}
\end{figure}

In Fig.~\ref{fig:analysis_cameras}a we show a typical distribution of input views in our manifold parameterization.
The distributions of multiple datasets can be found in our supplemental video.
We observe that our casual capture tends to result in an uneven view distribution and a significant fraction of the input views lying outside the manifold boundaries.
These views are vital for successful geometry reconstruction (Sec.~\ref{sec:calibration}), as they cover the subject from a wider range of directions, but these views cannot be used in the first training stage (Sec.~\ref{sec:training}).
We did not observe a strong effect of the exact view distribution when input views are reasonably stratified.

To investigate how the number of input views influences result quality, we progressively reduce the views used in our pipeline.
We then perform an exhaustive leave-one-out image quality analysis (following the same protocol as in Sec.~\ref{sec:comparisons}) for each configuration.
The results are given in Fig.~\ref{fig:analysis_cameras}b.
We observe that, unsurprisingly, image quality improves as the number of input views increases, but tends to saturate at about 20 views.
Using less than 15 views has a stronger negative effect on image quality.
For less than 10 views camera calibration and geometry reconstruction are unreliable.

\subsubsection{Camera Manifold}

\begin{figure}
	\includegraphics[width=0.99\linewidth]{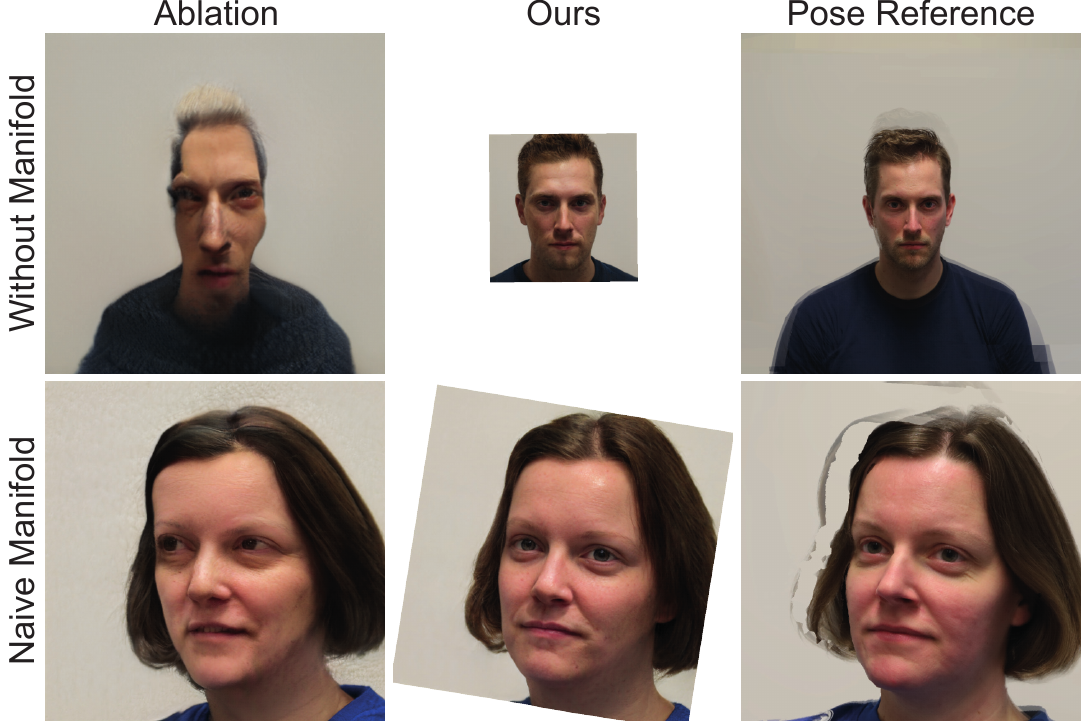}
	\caption{
		Manifold ablations:
		Training a mapping from free cameras to latents gives heavily distorted results (upper left).
		A naive mainfold is better, but still does not respect training data alignment and therefore results in distortions (lower left).
		In contrast, our solution (middle column) is distortion-free and matches the pose reference (right column) well.
		}
	\label{fig:ablation_manifold}
\end{figure}

We analyze three alternatives to our camera manifold formulation.
In the upper row of Fig.~\ref{fig:ablation_manifold}, we show a result obtained when omitting the manifold completely, \ie using the trackball camera model from Eq.~\ref{eq:trackballParam}.
We feed the four camera parameters to $M$ and train it to produce latents for free-viewpoint images directly.
As this task is harder, we give $M$ more capacity by doubling the number of both the hidden layers and feature channels.
The resulting images are severely distorted, while our approach matches the reference - a ULR rendering of the desired pose - well.

The lower row of Fig.~\ref{fig:ablation_manifold} shows a result obtained with a naive manifold:
We use the frontal pose to optimize for a coefficent vector $\hat{\textbf{c}}$ and fix it while rotating the camera around the head.
This naturally results in in-plane rotations and therefore out-of-distribution images for StyleGAN and we again observe strong distortions in the results, while our full approach handles these configurations well.

\begin{figure}
	\includegraphics[width=\linewidth]{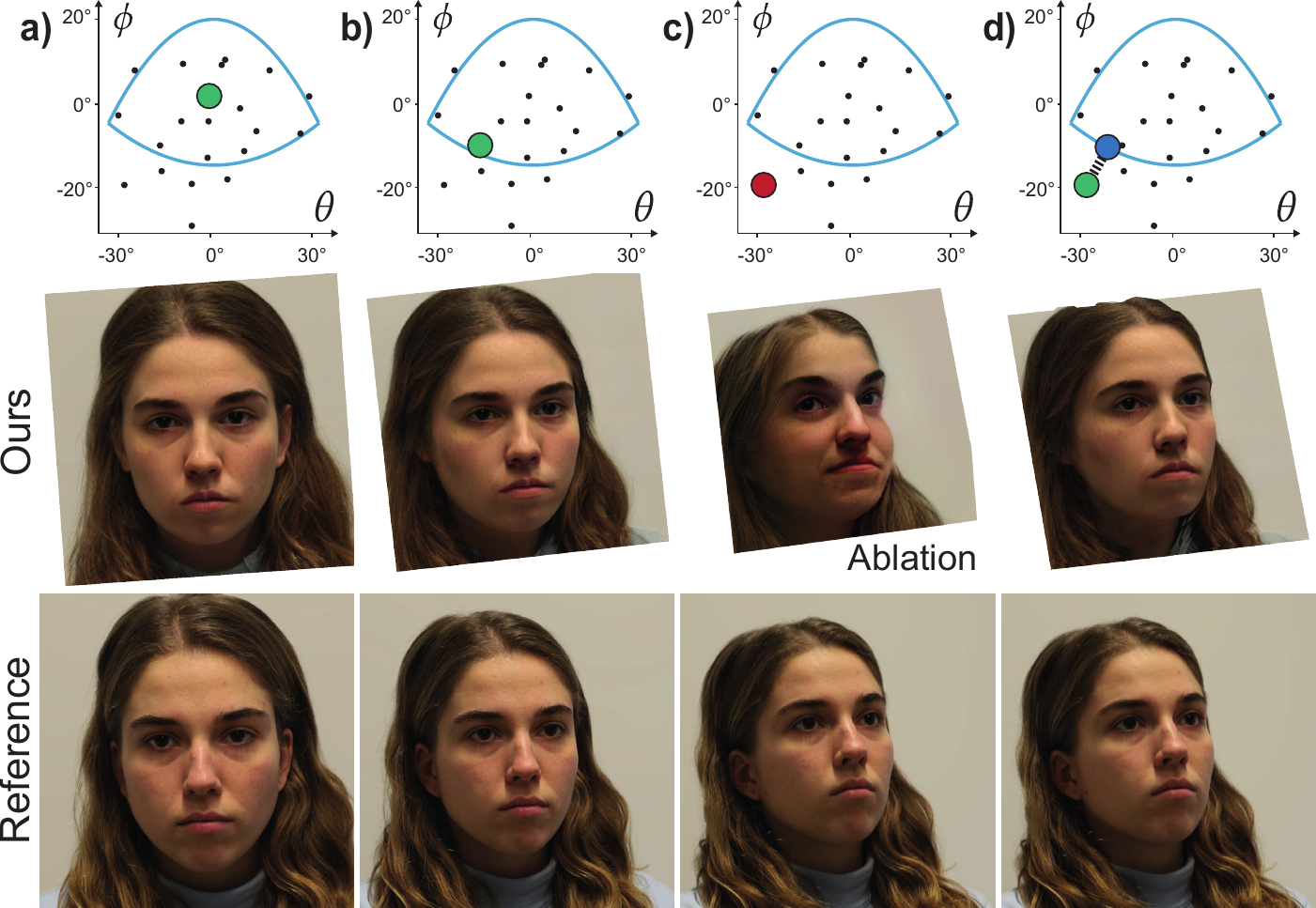}
	\vspace*{-2mm}
	\caption{
		Manifold boundary analysis:
		\emph{a}) and \emph{b}) Within the manifold boundaries our results are of consistently high quality.
		\emph{c}) Directly generating a view outside the manifold (red point) using StyleGAN results in low quality compared to the held-out input view. 
		\emph{d}) We first render the closest view in the valid range (blue point) with StyleGAN. The final image (green point) is then produced using warping. Black dots denote input views.
	\vspace*{-2mm}
}
	\label{fig:ablation_manifold_range}
\end{figure}

Finally, we analyze the effect of our manifold boundaries in Fig.~\ref{fig:ablation_manifold_range}.
Result quality is consistently high when moving within the boundaries (Fig.~\ref{fig:ablation_manifold_range}a and b; see also supplemental videos). 
When we omit the manifold boundaries and use $M$ to directly generate a view outside the valid range (Fig.~\ref{fig:ablation_manifold_range}c), image quality suffers.
Our full approach (Fig.~\ref{fig:ablation_manifold_range}d) first generates the closest valid StyleGAN image, and then warps it to the desired view, resulting in superior quality.

\subsubsection{Loss and Training}

\begin{figure}
	\includegraphics[width=0.99\linewidth]{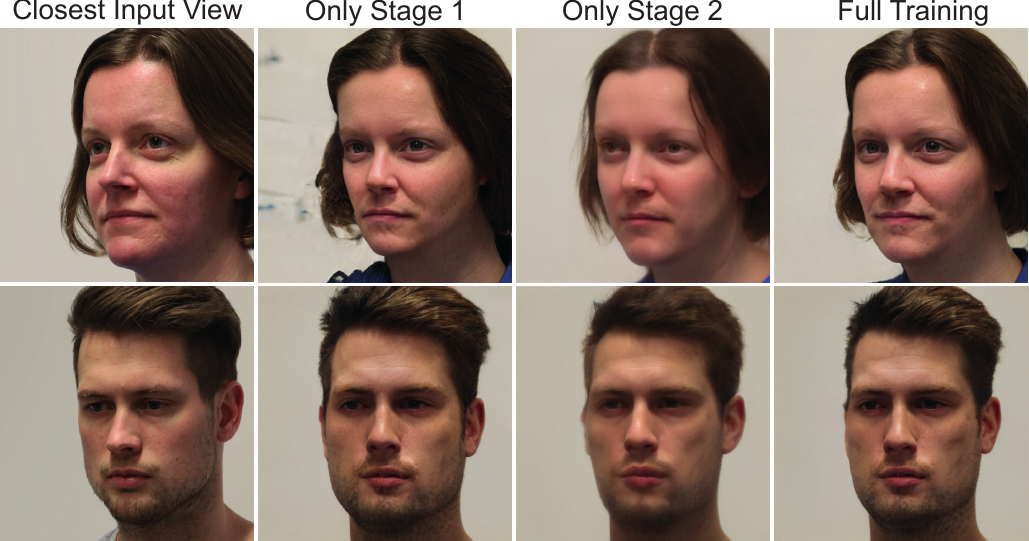}
	\vspace*{-2mm}
	\caption{
		Effect of the training stages:
		Only running stage 1 results in poor generalization to the entire manifold.
		Stage 2 alone gives overall lower quality results, as the high-quality information from the input views is missing.
		Our two-stage training procedure provides highest-quality results.
	\vspace*{-1mm}
}
	\label{fig:ablation_training_stages}
\end{figure}

We analyze our two-stage training procedure in Fig.~\ref{fig:ablation_training_stages}.
When only using the first stage for training, our method relies on the sparse input views only.
It therefore fails to generalize to the entire camera manifold, either by generating distorted images (first row) or false perspectives (second row).
If we only use the second training stage, we get blurry results with identity shifts due to IBR artifacts in the training data and the omitted LPIPS loss.

In the supplemental we also show that excluding the LPIPS term reduces image sharpness, while the identity loss preserves slight face identity shifts.
The prior loss increases photo-realism.

\subsubsection{Background Blur}

\begin{figure}
	\includegraphics[width=0.95\linewidth]{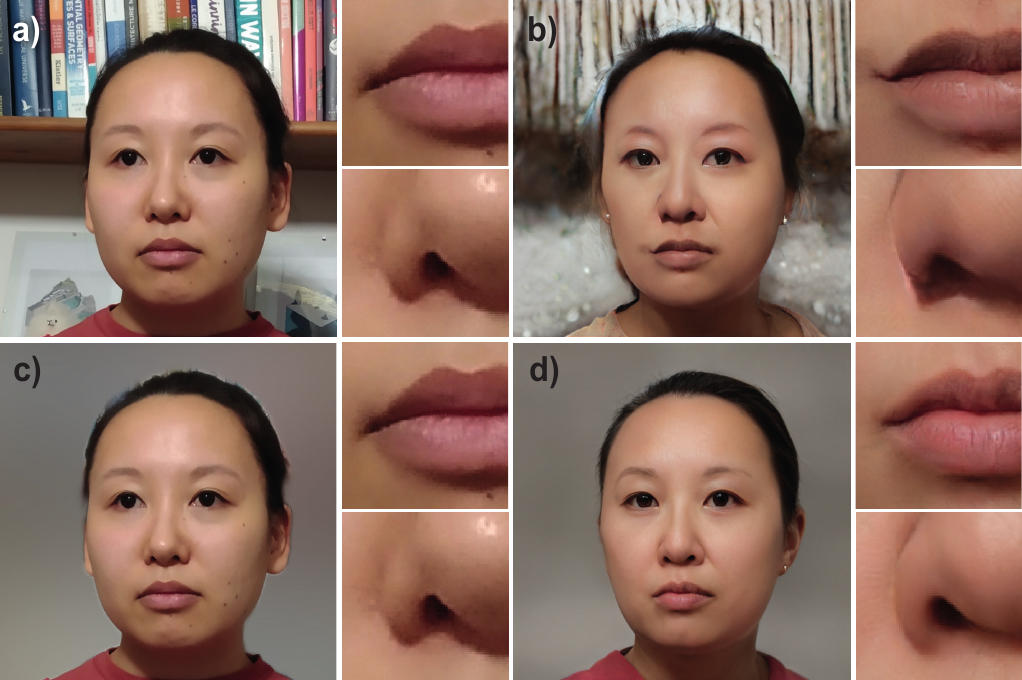}
	\vspace*{-2mm}
	\caption{
		If the background in the input images is not smooth (\emph{a}), the embedding does not only fail to match the background (\emph{b}), but also distorts facial features, such as mouth and nostrils (insets).
		We blur the background before the embedding (\emph{c}) to achieve higher-quality results (\emph{d}).
}
	\label{fig:ablation_background_blur}
	\vspace*{-2mm}
\end{figure}

Fig.~\ref{fig:ablation_background_blur} demonstrates the effect of blurring the background of the input images.
We observe that a blurred background has an influence not only on the background region, but also helps to preserve the identity and increase photorealism.

\section{Discussion, Future Work, and Conclusion}
\label{sec:discussion}

Our approach enables high-quality free-viewpoint synthesis of faces using StyleGAN with casually captured multi-view data as input.
While our capture is lightweight, previous work on GAN embeddings require only a single image; at the expense of significantly less variation in pose and no direct way to connect to precise camera parameters.
A possible extension of our method towards single-view inputs would require performing GAN inversion, geometry estimation, and camera calibration -- likely using a full camera model -- from a single image.
This could be done by harnessing the capabilities of StyleGAN as a multi-view generator \cite{zhang2021image} and to fold geometry estimation into the training loop \cite{pan20202d}.

\begin{figure}
	\includegraphics[width=0.66\linewidth]{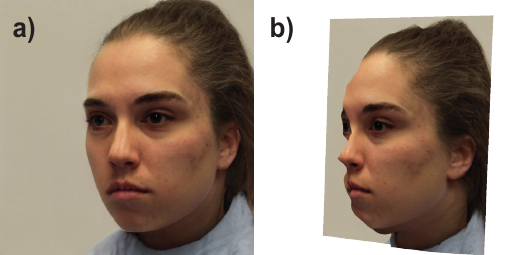}
	\vspace*{-2mm}
	\caption{
		Limitations:
		\emph{a}) If the camera position is in the valid manifold range, no parallax occurs and rendering quality is not impacted by geometric fidelity.
		\emph{b}) When leaving the valid manifold range, inaccuracies in the proxy geometry can become visible at extreme angles.
	\vspace*{-4mm}
	}
	\label{fig:limitation}
\end{figure}

We allow free positioning of the virtual camera, 
beyond the distribution of the StyleGAN training corpus.
If the camera position lies within the valid range of our camera manifold, the image quality of our method does not depend on the quality of the geometric reconstruction, as no parallax occurs (Fig.~\ref{fig:limitation}a);
if it leaves the valid range, our warping scheme generates the required parallax.
This can create artifacts if the proxy is incorrect or too coarse (Fig.~\ref{fig:limitation}b).
Further, all semantic manipulations that change geometric features of the face are naturally not mirrored in the geometry, leading to projective texturing artifacts.
A possible avenue of future work would be to synchronize geometric changes using a 3DMM.

StyleGAN embeddings need to find a trade-off between high-fidelity image reconstruction and high-quality editing capabilities \cite{tov2021designing, blau2018perception}.
Our approach seeks to find a sweet spot on this spectrum, but sometimes slight shifts in identity or other image attributes can occur.
A related issue is a ``showerdoor effect'', resulting in flickering and texture details sticking to the screen rather than the face.
We share these problems with previous work on this topic, as they stem to a large extent from inherent limitations of the generator \cite{Karras2021alias}.
In an orthogonal line of research, general embedding strategies have been explored, which do not require face-specific optimizations to obtain latent codes \cite{richardson2020encoding}. 
While these encoder-based approaches open up exciting research directions, the quality of the resulting embeddings is currently insufficient (see supplemental). 

It is potentially possible to train StyleGAN on a more diverse set of images such that free viewpoint capabilities naturally arise.
We believe that our approach of extending the capabilities of a constrained model in a post-process is a viable solution presenting a good compromise, given the resources needed to train a GAN from scratch, and the quality of current methods achieved by careful alignment.
An exciting avenue of future work would be to jointly train our embedding network and fine-tune the GAN to lift some of its limitations.

In conclusion, we have presented a method that allows the generation of a StyleGAN image from a free-viewpoint 3D camera, enabling these stunningly realistic images to be used in 3D applications, together with the semantic editing capabilities of previous methods. We believe that this is an important step forward in bridging the gap between powerful 2D image-processing learning solutions and the generation of fully editable 3D content for general use, with minimal content creation effort.


\section*{Acknowledgements}

This research was funded by the ERC Advanced grant FUNGRAPH No 788065 (\textcolor{blue}{\url{http://fungraph.inria.fr}}). The authors are grateful to the OPAL infrastructure from Université Côte d'Azur for providing resources and support. The authors thank Ayush Tewari, Ohad Fried, and Siddhant Prakash for help with comparisons, Adrien Bousseau, Ayush Tewari, Julien Philip, Miika Aittala, and Stavros Diolatzis for proofreading earlier drafts, the anonymous reviewers for their valuable feedback, and all participants who helped capture the face datasets.


\bibliographystyle{ACM-Reference-Format}
\bibliography{ms.bib}


\end{document}


\title{FreeStyleGAN: Free-view Editable Portrait Rendering with the Camera Manifold - Supplemental Materials}

\author{Thomas Leimk\"uhler}
\email{thomas.leimkuehler@mpi-inf.mpg.de}

\author{George Drettakis}
\affiliation{
	\institution{Universit\'{e} C\^{o}te d'Azur and Inria}
	\country{France}
}
\email{george.drettakis@inria.fr}

\maketitle

In this supplemental document we provide multi-view capture instructions, as well as methodology details on camera calibration and geometry reconstruction.
We also present specifics of the FFHQ alignment process, how to project manifold coordinates to the valid range and how to sample it, and our definition of a frontal pose.
Finally, we elaborate on our two-stage training procedure, provide additional comparisons and ablations, and show a comparison to an encoder-based embedding approach.

\section{Capture Instructions}

Here we provide the instructions we gave to the non-professional models and photographers who helped us with capturing the datasets.

\subsection{Instructions for the Model}

\begin{itemize}
\item You need to be as static as possible. Sit on a chair and use the backrest (make sure it is not visible above the shoulders).
\item Choose a point in front of you to look at during the entire session.
\item Make a neutral relaxed face, eyes open, mouth closed. No smiling please.
\end{itemize}

\subsection{Instructions for the Photographer}

\begin{figure}[!h]
	\includegraphics[width=0.7\linewidth]{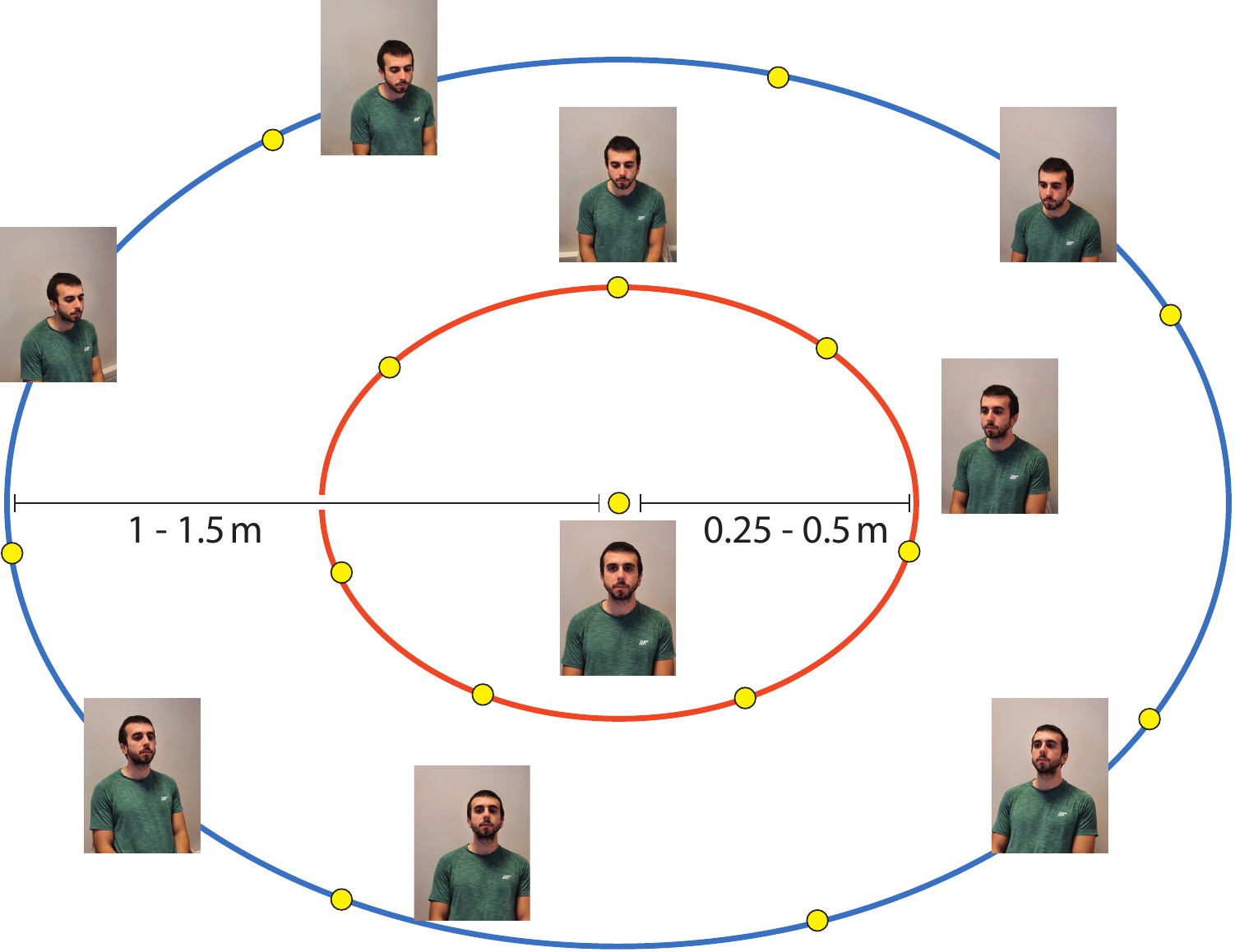}
	\vspace*{-3mm}
	\caption{
		Figure provided to the photographer to illustrate the capture distribution, encouraging a stratified set of camera poses. Yellow dots mark example capture positions.
	\vspace*{-2mm}
		}
	\label{fig:capture_instructions}
\end{figure}

\begin{itemize}
\item Don't use a flash. Avoid casting hard shadows onto the model with your body. Lighting should not be too harsh, shades on the face are fine and even appreciated.
\item Have a distance of about 1-2 meters to the model.
\item Take 10-25 pictures of the face and full upper body. Frame every view such that that top of the head is slightly below the upper image boundary and the belly button marks the lower image boundary. Please capture:
\begin{itemize}
\item One frontal view.
\item 5-10 views on an ellipse, about 0.25 to 0.5 meters around the frontal view (red in Fig.\ref{fig:capture_instructions}).
\item 5-10 views on an ellipse, about 1.0 to 1.5 meters around the frontal view (blue in Fig.\ref{fig:capture_instructions}).
\item Add more views at will.
\end{itemize}
\item Capture positions don't have to be exact.
\end{itemize}

\noindent
All results in the paper were produced using our own captures, except for the first row in Fig. 10, which uses material from Milborrow \etal \shortcite{Milborrow10}.

\section{3D Calibration and Reconstruction}
We obtain 
camera calibration and the geometric proxy using off-the-shelf software \cite{realitycapture}. First, the virtual input cameras are calibrated using structure from motion \cite{snavely2006photo}, then a multi-view stereo algorithm estimates the 3D shape using dense pixel correspondences \cite{goesele2007multi, realitycapture}, followed by a meshing step to obtain a triangle mesh of the face. 
We 
smooth the mesh \cite{sorkine2005laplacian} to get rid of high-frequency reconstruction noise.
The output of this process are the reconstructed triangle mesh, the calibrated cameras and possibly resampled input images with a lens distortion correction applied.
Even though more than 10-25 cameras are usually recommented for 3D reconstruction, quality is high enough for our method, despite taking the photos casually without a rig.

\section{Review of FFHQ Alignment}
\label{sec:ffhq_alignment}

\begin{figure}[!h]
	\includegraphics[width=0.99\linewidth]{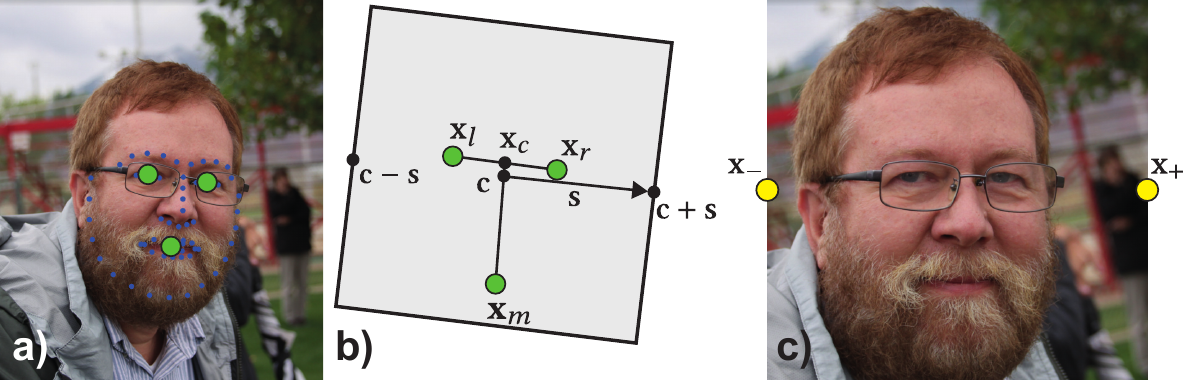}
	\vspace*{-2mm}
	\caption{
		The 2D alignment performed in the FFHQ dataset.
		\emph{a)} Raw facial feature points (blue dots) are detected and aggregated to obtain representative eye and mouth positions (green dots).
		\emph{b)} Geometric features are used to determine the square crop window (grey, not shown to scale) with center $\mathbf{c}$ and vector $\mathbf{s}$ giving orientation and size.		
		\emph{c)} The resulting aligned image.
	\vspace*{-2mm}
		}
	\label{fig:ffhq_alignment}
\end{figure}

The FFHQ dataset was constructed by first collecting images from Flickr, followed by a cleanup step and alignment, where first 68 facial features are found \cite{kazemi2014one} (blue dots in Fig.~\ref{fig:ffhq_alignment}a). 
Then the eye and mouth features are aggregated to obtain representative eye positions $\mathbf{x}_l$ and $\mathbf{x}_r$, as well as a mouth position $\mathbf{x}_m$ in the image (green dots in Fig.~\ref{fig:ffhq_alignment}a).
From these three points a crop window is computed as follows (Fig.~\ref{fig:ffhq_alignment}b):
The center of the window is computed as a convex combination of the eye midpoint $\mathbf{x}_c = 0.5 \left(\mathbf{x}_l + \mathbf{x}_r \right)$ and the mouth, as in
$\mathbf{c} = \lambda \mathbf{x}_c + (1 - \lambda) \mathbf{x}_m$,
where $\lambda=0.9$ in the reference implementation.
The eye and mouth positions are also used to determine the orientation of the square crop window:
\begin{equation*}
\hat{\mathbf{s}}
=
\mathbf{x}_r - \mathbf{x}_l + \text{rot}_{90\degree}(\textbf{x}_m - \mathbf{x}_c)
\end{equation*}
is used as the horizontal orientation of the crop window, where $\text{rot}_{90\degree}$ denotes a counter-clockwise rotation of $90\degree$.
The size of the crop window should encompass the entire head, and the heuristic approach in the reference implementation reads as
\begin{equation*}
\mathbf{s}
=
\max
\left(
2 \| \mathbf{x}_l - \mathbf{x}_r \|,
1.8 \| \mathbf{x}_m - \mathbf{x}_c \|
\right)
\frac
{\hat{\mathbf{s}}}
{\| \hat{\mathbf{s}} \|}.
\end{equation*}
Given the above crop window geometry, the original image is re-sampled to obtain the final aligned output image (Fig.~\ref{fig:ffhq_alignment}c).

\section{Manifold Range Projection}
\label{sec:appendix_range_projection}

Given the rotational components of a manifold coordinate
$q= \left[\theta, \phi \right]^T$,
the closest point $q^*$ in the valid region of the manifold is
\[
q^*
=
\begin{cases} 
q & \text{if } c_l(\theta) \leq \phi \leq c_u(\theta)\\ 
\left[ g, c_u(g) \right]^T & \text{if } c_l(\theta) > \phi > c_u(\theta)\\
\left[ h_u, c_u(h_u) \right]^T & \text{if } c_l(\theta) < c_u(\theta) < \phi \\ 
\left[ h_l, c_l(h_l) \right]^T & \text{if } c_u(\theta) > c_l(\theta) > \phi 
\end{cases}
\]
where
\begin{equation*}
\begin{split}
g 
&= 
\text{sign}(\theta)\sqrt{\frac{b_l-b_u}{a_u-a_l}},
\\
h_i 
& = 
\sqrt[3]{\frac{\theta}{4a_i^2}+D_i} + \sqrt[3]{\frac{\theta}{4a_i^2}-D_i},
\quad
\text{and}
\\
D_i 
& = 
\sqrt{\left( \frac{1+2a_i(b_i-\phi)}{6a_i^2} \right)^3 + \left(\frac{\theta}{4a_i^2}\right)^2},
\end{split}
\end{equation*}
with 
$c_i(\theta) = a_i \theta^2 + b_i$
and
$i \in \{ u,l \}$.


\section{Manifold Sampling}
\label{sec:appendix_sampling}

To sample the valid range of the camera manifold during training, we employ the inverse-CDF method.
We observe that $p(\theta)$ is proportional to the difference of the two bounding parabolas:
\[
p(\theta)
=
\frac{1}{Z}
\left(
c_u(\theta) - c_l(\theta)
\right)
=
\frac{1}{Z}
\left(
\Delta a \theta^2 + \Delta b
\right),
\]
where
$\Delta a = a_u - a_l$
and
$\Delta b = b_u - b_l$.
Further, $\theta$ is defined between the intersections of the two bounding parabolas, \ie
$\theta \in \left[-g, g \right]$,
where
$g = \sqrt{-\frac{\Delta b}{\Delta a}}$.
Therefore,
\[
Z =
\int_{-g}^{g}
\left(
\Delta a \theta^2 + \Delta b
\right)
\mathrm{d} \theta
=
2
\left(
\frac{\Delta a}{3}
g^3
+
\Delta b
g
\right).
\]
The CDF is given by
\[
P(\theta)
=
\int_{-g}^\theta
p(\theta')
\mathrm{d} \theta'
=
\frac{1}{Z}
\left(
\frac{\Delta a}{3}
\theta^3
+
\Delta b \theta
\right)
+ \frac{1}{2}.
\]
To obtain a valid manifold sample
$\left[ \theta, \phi, d \right]^T$, 
we draw three canonical uniform random samples $\boldsymbol{\xi}$.
Then
$
\theta 
= 
P^{-1}
\left(
\xi_1
\right)
$,
which we evaluate numerically, and
$
\phi
=
\xi_2 
\left( 
c_l(\theta) - c_u(\theta)
\right)
+ 
c_l(\theta).
$
Finally, we linearly remap $\xi_3$ to obtain $d$.

\section{Frontal Pose Calibration}
\label{sec:feature_conversion}

For the definition of the manifold (Sec. 4.1) and its range (Sec. 4.2), as well as the 3D alignment of the face mesh (Sec. 4.4), we need to consistently define how 3D eye and mouth positions are related to manifold coordinates $\theta$ and $\phi$. 
For calibration, we consider the frontal pose $\theta = \phi = 0$.
The horizontal orientation is straightforward: 
We set $\theta=0$ when both eye locations have the same depth.
For the vertical orientation, there exists no obvious frontal pose.
We therefore set $\phi=0$ when the depth of the mouth is one eighth of the interocular distance smaller than the depth of the eye midpoint.
This configuration is somewhat arbitrary and could be replaced by any suitable alternative.
Note, however, that the exact orientation does not change results as long as we are consistent with these definitions.


\section{Details on the Training Procedure}
We found a progressive training schedule, which splits training into two stages, to produce results of highest quality.
\begin{figure}
	\includegraphics[width=0.99\linewidth]{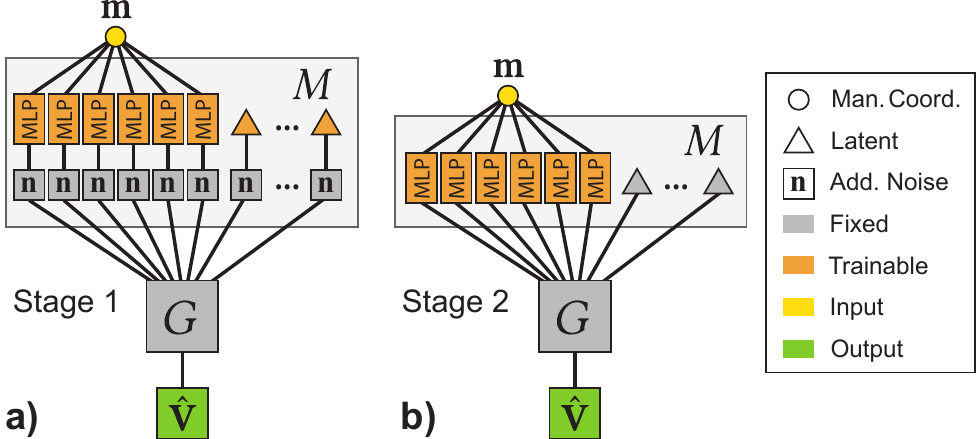}
	\vspace*{-2mm}
	\caption{
	We employ a progressive training schedule.
		\emph{a)} In the first stage, we only use the input images as training data. We train MLPs that map manifold coordinates (yellow dot) to the first 6 StyleGAN latents, and we directly optimize for the remaining static latents (orange triangles). All latents fed to $G$ are subject to random perturbations (boxes labelled $\mathbf{n}$) during training.
		\emph{b)} In the second stage, we fine-tune the MLPs by augmenting the training data with IBR and fix the static latents (grey triangles).
		}
	\label{fig:training}
\end{figure}
\subsection{Stage 1: Sparse Input Views}
\label{sec:training_stage1}

In the first stage, we only use the aligned input views as training data and optimize all trainable parameters (Fig.~\ref{fig:training}a).
Following Karras \etal \shortcite{karras2020analyzing}, we initialize all latents with
$
\mathbf{\mu}_\mathbf{w}
=
\mathbb{E}_\mathbf{z}
H(\mathbf{z})
$.
This initializes the optimization with the ``average face'' in latent space $W$ and is obtained by running 1000 random codes $\mathbf{z}$ through the mapping network $H$.
We run our optimization for 7500 iterations using Adam \cite{kingma2014adam} with default parameters and a batch size of 2.
We start with a learning rate of $0.005$ and decay it exponentially by a factor of 0.98 every 200 iterations.
Again following Karras \etal \shortcite{karras2020analyzing}, we allow the optimization to escape local minima by adding stochastic Gaussian random noise to the latents in each training iteration.
The optimization starts with a noise standard deviation of $\sigma=0.1$ and decays as described in Karras \etal
We use the following weights for our loss terms:
$\lambda_\text{LPIPS}=100$, and
$\lambda_\text{id}=1$.
Following ideas from Tewari \etal \shortcite{tewari2020pie}, we vary $\lambda_\text{prior}$ over time:
For the first 2500 iterations we set $\lambda_\text{prior}=10$ to ensure a reasonable embedding close to $W$.
For the remaining iterations we set $\lambda_\text{prior}=0.1$, which allows the optimization to explore the extended space $W^+$.
Intuitively, this training stage provides sparse anchors for the MLP, which is responsible for pose changes and at the same time optimizes the latents of the static GAN layers with the highest-possible quality training data.
Training this stage takes 35 minutes on an NVIDIA RTX6000.
\subsection{Stage 2: Dense Manifold}

In the second stage, we provide samples from the entire manifold as training data using a mixture of ULR renderings ($85\%$) and input views ($15\%$).
Now we fix the latents of the static detail layers to prevent high-frequency IBR artifacts from impacting them (Fig.~\ref{fig:training}b).
Only the MLP weights are refined at this stage, essentially filling in the pose gaps between the input views.
We run this stage for 750 iterations with the same optimization parameters as before, but omitting the random noise and using the following (constant) weights for our loss terms:
$\lambda_\text{LPIPS}=0$,
$\lambda_\text{id}=1.5$, and
$\lambda_\text{prior}=0.1$.
Disabling the LPIPS loss term is motivated by the fact that it is very sensitive to IBR artifacts and no obvious way exists to incorporate uncertainty akin to Eq.~6 into the multiscale VGG architecture \cite{Liu_2018_ECCV} without altering its perceptual prediction quality.
However, we found $\mathcal{L}_{\ell_1}$ as the sole image quality loss sufficient for this fine-tuning stage.
Training this stage takes 4 minutes.

\begin{figure}
	\includegraphics[width=0.99\linewidth]{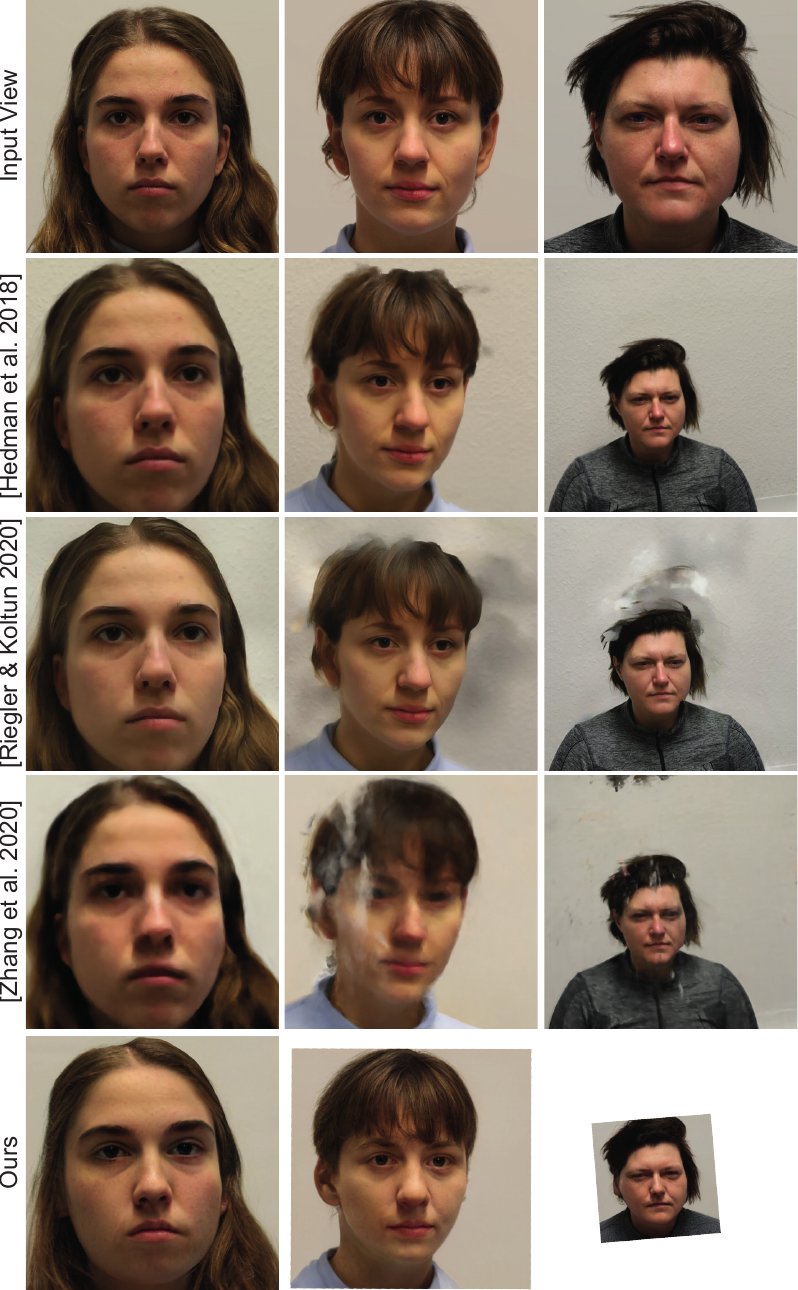}
	\vspace*{-3mm}
	\caption{
		Comparisons to free-viewpoint rendering methods.}
	\label{fig:comparisons_free_viewpoint}
	\vspace*{-5mm}
\end{figure}

\begin{table*}[]
\begin{threeparttable}
	\centering
	\caption{
	Comparing camera accuracy (measured using facial landmarks), image quality, and face recognition error against free-view rendering methods.}
	\vspace*{-4mm}
	\label{tab:quant_free}
	\begin{tabular}{lcrrrrrr}
		Method & Semantic Editing & \multicolumn{2}{c}{Facial Landmarks} & \multicolumn{3}{c}{Image Quality} & Recognition Error$\downarrow$ \\		
		 & & Alignment$\downarrow$ & Detection Rate$\uparrow$ & PSNR$\uparrow$ & SSIM$\uparrow$ & E-LPIPS$\downarrow$ & \\
		\toprule
		\cite{hedman2018deep} & \xmark & .023 & 99\% & 26.4 & .875 & .010 & .07 \\
		\cite{riegler2020free} & \xmark & .027 & 100\% & 23.6 & .845 & .013 & .08 \\
		\cite{zhang2020nerf}\tnote{1} & \xmark & .018 & 100\% & 30.6 & .853 & .015 & .24   \\
		\cite{siarohin2020first} & \cmark\tnote{2} & .254 & 42\% & 11.7 & .531 & .061  & .23  \\				
		Ours & \cmark & .068 & 100\% & 20.9 & .717 & .017  & .14  \\
		\bottomrule
	\end{tabular}
\begin{tablenotes}\footnotesize
\item[1] Due to time constraints, we did not train a separate model for each leave-one-out image set, but only one model using all images per subject.
\item[2] Editing is restricted to facial animations using a driving video.
\end{tablenotes}
\end{threeparttable}
\vspace*{-2mm}
\end{table*}

\section{Additional Comparisons}

In addition to the comparisons shown in the main paper, we present more details on the comparisons to other image-based rendering methods in Fig.~\ref{fig:comparisons_free_viewpoint} and Tbl.~\ref{tab:quant_free}.
We compute image quality using the PSNR, SSIM \cite{wang2004image}, and E-LPIPS \cite{kettunen2019lpips} metrics of the facial region.
We see that - not surprisingly - the neural IBR methods win most of the competitions. This comes at the cost of static scenes, which cannot be edited. 
Due to time constraints, we did not train a separate NeRF++ model for each leave-one-out image set.
The numerical results on facial landmarks and image quality are therefore heavily skewed in favor of the method.
Please see the supplemental video for novel-view camera paths.

The method of Siarohin et~al., which allows semantic editing in the form of facial expressions, does not perform well in the free-viewpoint setting for the metrics we considered.
We observe that our approach allows free-view synthesis with camera accuracy in the order of magnitude of the IBR methods, while obtaining decent image quality and facial identity scores -- while inheriting the full editing potential of StyleGAN.
Note that the comparison to IBR is only possible because our method for the first time allows precice camera control of GAN imagery.

\section{Loss Term Ablations}

\begin{figure*}
	\includegraphics[width=0.85\linewidth]{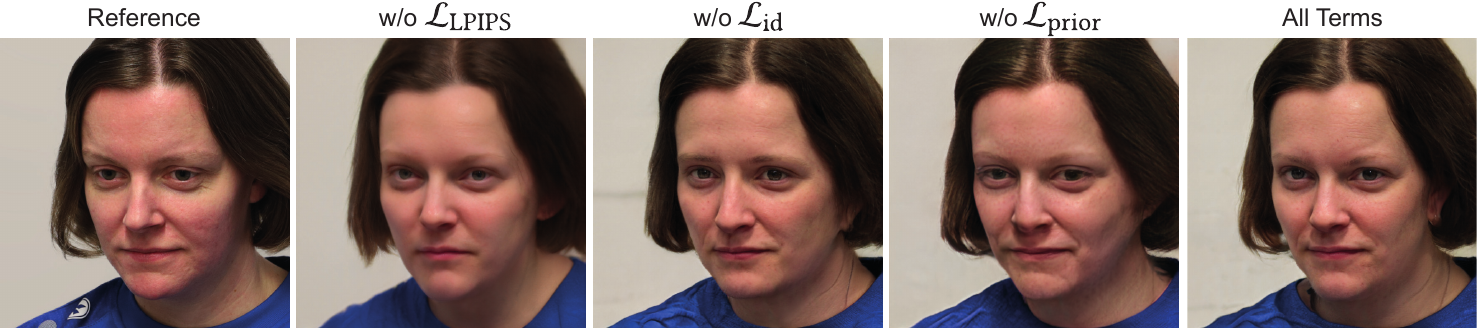}
	\vspace*{-3mm}
	\caption{
		Loss term ablation. Only our full loss formulation gives best results.
}
	\label{fig:ablation_losses}
\end{figure*}

Fig.~\ref{fig:ablation_losses} shows the effect of our individual loss terms.
Excluding the LPIPS term reduces image sharpness, while the identity loss preserves slight face identity shifts.
The prior loss increases photo-realism.
We observe that all of our loss terms are necessary to produce results of high fidelity.

\begin{figure}
	\includegraphics[width=0.92\linewidth]{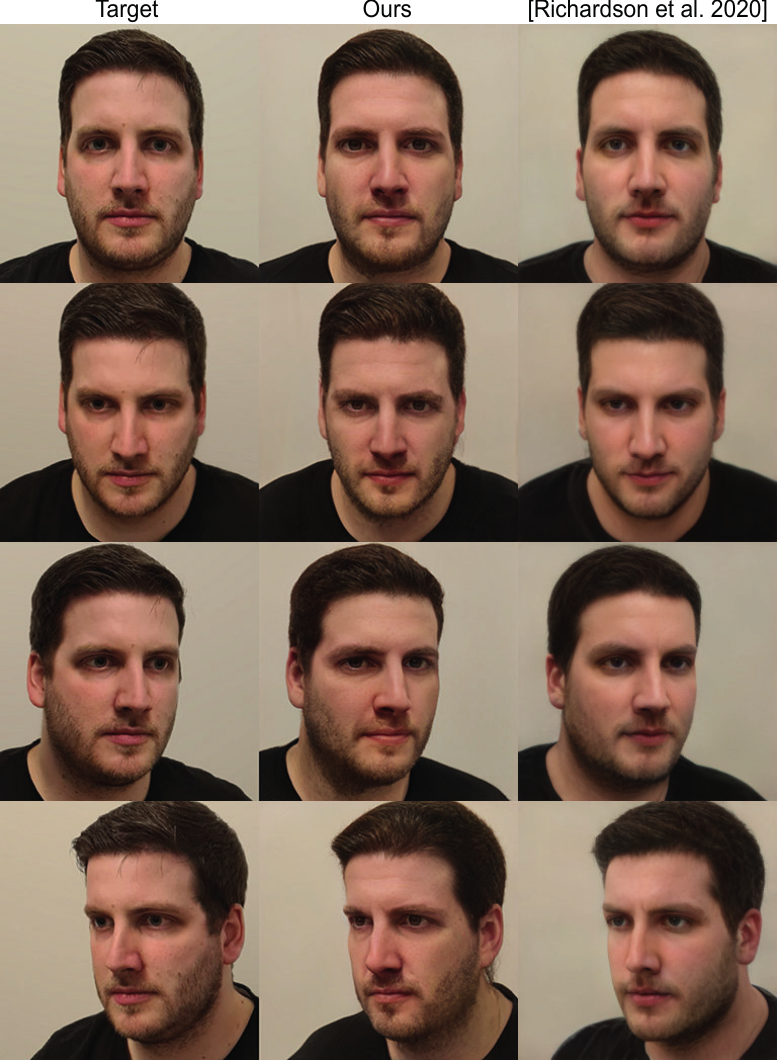}
	\vspace*{-2mm}
	\caption{
		Comparing our optimization-based embedding strategy (center) with the state-of-the-art encoder-based method of Richardson \etal \shortcite{richardson2020encoding} (right). We observe that our approach produces a more faithful reconstruction of the target (left).
}
	\label{fig:comparison_pspnet}
\end{figure}

\section{Encoder-based Embeddings}

General embedding strategies have been explored, which do not require face-specific optimizations to obtain latent codes \cite{richardson2020encoding}. 
While these encoder-based approaches open up exciting research directions, the quality of the resulting embeddings is currently insufficient (see Fig.~\ref{fig:comparison_pspnet}).


\bibliographystyle{ACM-Reference-Format}
\bibliography{ms.bib}
